\documentclass[12pt]{article}

\usepackage{latexsym,amsmath,amscd,amssymb,amsthm,graphics}
\usepackage{enumerate}
\usepackage[margin=2cm]{geometry}

\usepackage{graphicx}

\usepackage[square,authoryear]{natbib}
\usepackage{url}
\usepackage{framed}
\usepackage[all]{xy}
\usepackage{epstopdf}

\makeatletter

\@addtoreset{figure}{section}
\def\thefigure{\thesection.\@arabic\c@figure}
\def\fps@figure{h, t}
\@addtoreset{table}{bsection}
\def\thetable{\thesection.\@arabic\c@table}
\def\fps@table{h, t}
\@addtoreset{equation}{section}

\makeatother

\begin{document}

\newtheorem{theorem}{Theorem}[section]
\newtheorem{definition}[theorem]{Definition}
\newtheorem{lemma}[theorem]{Lemma}
\newtheorem{remark}[theorem]{Remark}
\newtheorem{proposition}[theorem]{Proposition}
\newtheorem{corollary}[theorem]{Corollary}
\newtheorem{example}[theorem]{Example}

\def\balpha{\boldsymbol\alpha}
\def\bom{\boldsymbol\omega}
\def\bOm{\boldsymbol\Omega}
\def\bPi{\boldsymbol\Pi}
\def\below#1#2{\mathrel{\mathop{#1}\limits_{#2}}}



\title{Stochastic geometric models with non-stationary spatial correlations in Lagrangian fluid flows}

\author{Fran\c{c}ois Gay-Balmaz$^1$ and Darryl D Holm$^2$
}
\addtocounter{footnote}{1}
\footnotetext{CNRS and \'Ecole Normale Sup\'erieure de Paris, Laboratoire de M\'et\'eorologie Dynamique, 24 Rue Lhomond, 75005 Paris, France. 
\texttt{gaybalma@lmd.ens.fr}
\addtocounter{footnote}{1} }
\footnotetext{Department of Mathematics, Imperial College, London SW7 2AZ, UK. 
\texttt{d.holm@ic.ac.uk}
\addtocounter{footnote}{1}}

\date{}

\maketitle

\makeatother

\begin{abstract}
Inspired by spatiotemporal observations from satellites of the trajectories of objects drifting near the surface of the ocean in the National Oceanic and Atmospheric Administration's ``Global Drifter Program'', this paper develops data-driven stochastic models of geophysical fluid dynamics (GFD) with non-stationary spatial correlations representing the dynamical behaviour of oceanic currents. Three models are considered. Model 1 from Holm [2015] is reviewed, in which the spatial correlations are time independent. Two new models, called Model 2 and Model 3, introduce two different symmetry breaking mechanisms by which the spatial correlations may be advected by the flow. These models are derived using reduction by symmetry of stochastic variational principles, leading to stochastic Hamiltonian systems, whose momentum maps, conservation laws and Lie-Poisson bracket structures are used in developing the new stochastic Hamiltonian models of GFD. 
\end{abstract}

\maketitle

\newpage

\tableofcontents

\section{Introduction} \label{Intro-sec}

This paper develops data-driven stochastic models of fluid dynamics, inspired by spatiotemporal observations from satellites of the spatial paths of objects drifting near the surface of the ocean in the National Oceanic and Atmospheric Administration's ``Global Drifter Program''.  
The Lagrangian paths of these freely drifting instruments track the ocean currents. That is, the satellite readings of their positions approximate the motion of a fluid parcel as a curve parameterised by time. 

\begin{figure}[h!]
    \centering
    {\includegraphics[scale=0.5]{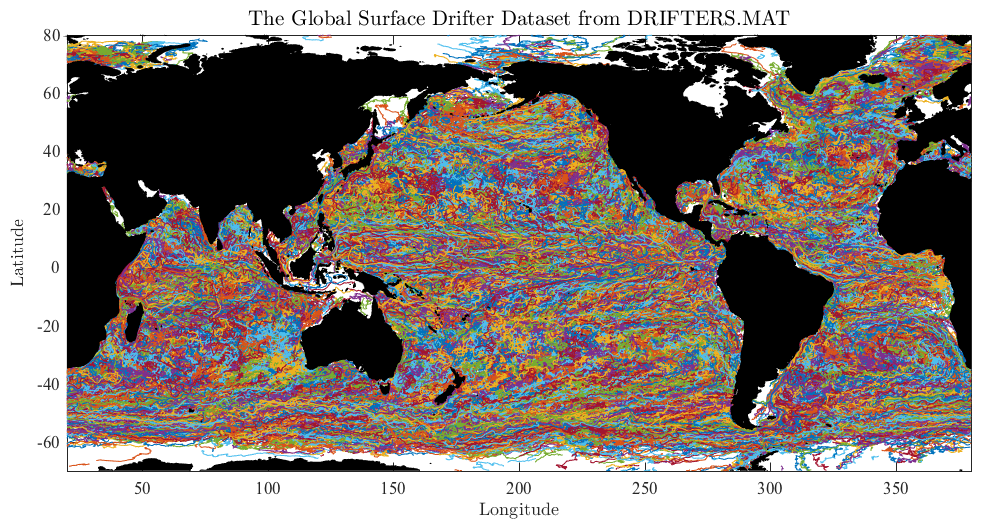} \label{fig:globaldrift}}
    \caption{Trajectories from the National Oceanic and Atmospheric Administration Global Drifter Program are shown, in which each colour corresponds to a different drifter.}
\end{figure}

\begin{figure}[h!]
    \centering
    {\includegraphics[scale=0.5]{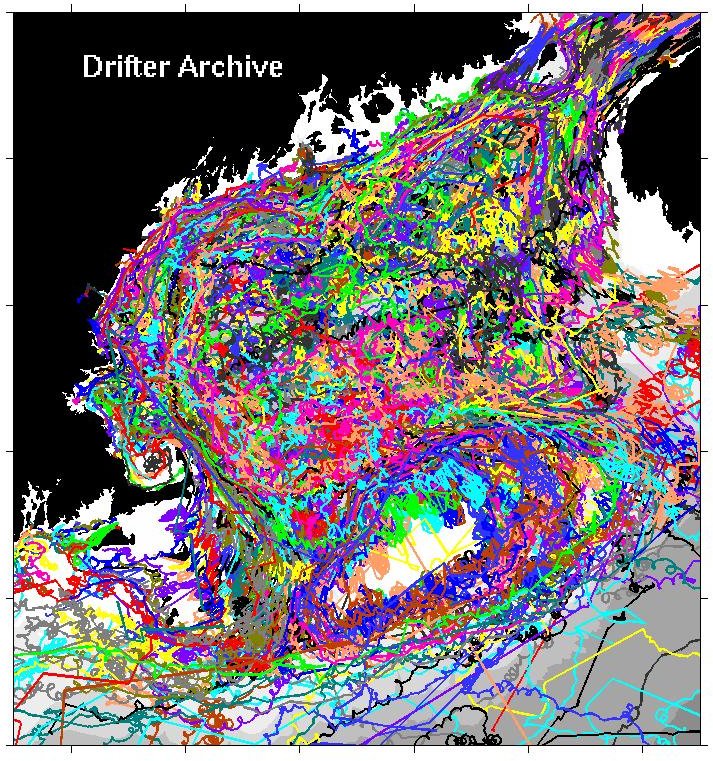} \label{fig:studentdrifters}}
    \caption{A subset of the drifter trajectories in the vicinity of Cape Cod.}
\end{figure}\vspace{-2mm}

\begin{figure}[h!]
\centering
    {\includegraphics[scale=0.5]{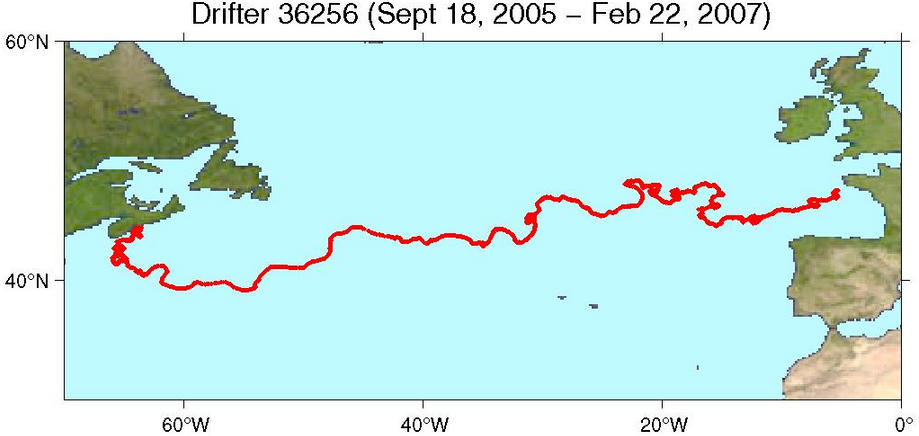} \label{fig:HT-drifter}}
    	\caption{The 521-day trajectory of North Atlantic drifter 36256. One notes the flow dependent interactions between fluid motions of several size scales. {Typical excursions of the drifters are about 10 kilometres or less, unless the drifter becomes entrained into an eddy.}
 }
\end{figure}$\,$

Figure 1.1 \cite{Lilly2017}  
displays the global array of surface drifter trajectories from the National Oceanic and Atmospheric Administration's ``Global Drifter Program'' (\url{www.aoml.noaa.gov/phod/dac}). In total, more than 10,000 drifters have been deployed since 1979, representing nearly 30 million data points of positions along the Lagrangian paths of the drifters at six-hour intervals.  This large spatiotemporal data set is a major source of information regarding ocean circulation, which in turn is an important component of the global climate system; see for example \cite{LuPa2007,Griffa2007,Sy-etal2016}. An important feature of this data is that the ocean currents show up as time-varying spatial correlations, easily recognised visually by the concentrations of colours representing individual paths. These spatial correlations exhibit a variety of spatial scales for the trajectories of the drifters, corresponding to the variety of spatiotemporal scales in the evolution of the ocean currents which transport the drifters.

Figure 1.2 
shows a sample of the Lagrangian trajectories of drifters released in the vicinity of Cape Cod. Here one sees concentrations of recirculating drifters following Western boundary currents, and splitting into three main streams just off the coast of Cape Cod. One of these main streams forms a large-scale re-circulation some distance away from the boundary and to the East of Cape Cod. Small-scale, erratic deviations of the main streams of the drifter paths are also visible in this Figure. These erratic trajectories will be represented stochastically in the paper, and the larger scale spatial correlations they follow will be modelled as spatiotemporal modulations of the stochasticity, which follow the resolved drift currents.

Figure 1.3 
displays the Lagrangian path taken by Drifter 36256, deployed on Sep 18, 2005, and  successfully recovered on Feb 21, 2007, at Brest, France, after a journey of 521 days across the North Atlantic Ocean. Along this path one sees the effects of the interactions of the drifter with a variety of space and time scales of of evolving fluid motion, strongly suggesting a need for modelling the non-stationary statistics of Lagrangian paths \cite{Sy-etal2016}.

\paragraph{Why introduce stochasticity into fluid dynamics?} In developing parameterisations in weather and climate prediction, the term ``unresolved scales'' refers to those fluid motions and thermodynamic properties which are either not computable in a given set of numerical simulations, or are not measurable in a given set of observations. The effects of the smaller, faster, unresolved scales on the larger, slower, resolved variables are often modelled deterministically, in terms of the resolved scales and their gradients. See, e.g., \cite{FoHoTi2001,FoHoTi2002,Ho2002} and references therein.
Although this deterministic approach may be pragmatic and even advisable, due to the limitations of simulations and measurements, it is clear from the variety of deterministic models which have been proposed that these models are by no means unique. For example, the plethora of turbulence models and averaging methods, as well as the stark differences in solution behaviour as simulations of the fluid equations are achieved at finer and finer resolutions, show that the resolved states may be associated with many possible unresolved states. For example,  phenomena such as vortex tubes of a certain radius, as seen in simulations at one scale, or resolution, may no longer even be present when simulated at smaller scales, achieved with better resolution. This is not the usual situation in the development of science. Usually, the advent of better, more accurate, measurements lead to new effects and new laws governing them. Instead, in the case of weather and climate prediction, the fundamental deterministic equations and the laws of thermodynamics are completely known. 

No new deterministic laws of fluid motion should be expected. However, there can be new \emph{statistical approaches} to the physical and mathematical descriptions of weather and climate. For example, the recognition of stochastic vector fields as the basic paradigm in fluid dynamics has long been the province of turbulence modelling \cite{MoninYaglom}. However, recently the use of stochastic vector fields has also been recognized in estimating statistical model uncertainty in numerical weather prediction, \cite{FrO'KBeWiLu2015}. From this viewpoint, the uncertainty and variability of the predictions are crucial aspects of the solution. In the statistical science of numerical weather and climate prediction, stochastic methods offer systematic approaches toward quantitative estimates of uncertainties due to model error and inaccuracy of data assimilation, as well as improved estimates of long-term climate variability, including estimates of the probability of extreme events. Following \cite{BeJuPa2013} we agree that ``stochasticity must be incorporated at a very basic level within the design of physical process parameterizations and improvements to the dynamical core.'' The purpose of this paper is to offer new approaches at this basic level. { However, it is beyond the scope of the present paper to analyse the large dataset of drifter trajectories that has inspired our investigation. The work is in progress to fully analyse the drifter dataset and will be discussed elsewhere. Here we will discuss both 2D and 3D stochastic fluid dynamics models.}

\paragraph{How to do it?} 
How does one use stochasticity to improve the physical and mathematical basis for designing statistical model uncertainty schemes? 
{ Recently, this question was answered by using Hamilton's principle to derive a new class of mathematical models of stochastic transport in fluid dynamics \cite{Ho2015}. In this class of models, the effects of the small, fast, unresolved fine scales of motion on the coarser ones are modelled by introducing stochastic uncertainty into the transport velocity of fluid parcels in the dynamics at the \textit{resolvable coarse scale}. This stochastic transport velocity decomposition encompasses both Newtonian and variational perspectives of mechanics and also leads to proper Kelvin circulation dynamics. For a rigorous derivation of the same decomposition into mean and fluctuating velocities using multi-time homogenisation methods, see \cite{CoGoHo2017}.  Applying this decomposition to create stochastic fluid dynamics  preserves the fundamental mathematical properties of their deterministic counterparts \cite{CrFlHo2017}. It also enables new approaches to sub-grid scale parameterization, expressed both in terms of fluctuation distributions, and spatial/temporal correlations. As such, it introduces stochastic corrections that are amenable to statistical inference from high-resolution data (either observed, or numerical). Moreover, this new class of models forms the ideal approach for the development of a novel data assimilation technology based on particle filters \cite{Beskos2017}. Filtering and ensemble techniques require de facto a stochastic representation of the dynamics. This randomization is most often achieved through random perturbations of the initial conditions. However, this approach tends to yield insufficient spreading of the ensemble and produce a poor representation of the error dynamics \cite{BeJuPa2013}. The stochastic transport class of models establishes the much-needed randomization via its rigorous derivation at the fundamental level, rather than via ad hoc empiricism. 

In this paper, we will introduce two different stochastic extensions of \cite{Ho2015} in applying the geometric mechanics framework to estimate the contribution of stochastic transport to statistical uncertainty (error) in fluid dynamics models. 
}

Stochastic transport means that the Lagrangian fluid parcel motion has a stochastic component. In this context, we ask the following question in the context of numerical weather and climate prediction: What fundamental properties of the deterministic fluid equations would persist in a stochastic vector field representation of continuum fluid motions? First, even if the fluid parcel velocity were stochastic, the fluid continuum motion would still be describable as a spatially smooth but now temporally stochastic flow, $g_t$, depending on time $t$.
For the sake of brevity and simplicity of the presentation, we will take the domain of flow $\mathcal{D}$ to lie in $ \mathbb{R}^2$ or $ \mathbb{R}^3$, and we will neglect considerations of boundary conditions in the examples discussed in this paper. Even in the presence of stochasticity, the stochastic path of the fluid parcel which is initially at position $X$ in the domain of flow
can still be represented by the formula for the Lagrange-to-Euler flow map, $x_t = g_t(X)$, so that $g_0$ at time $t=0$ is the identity map, $g_0(X)=X$. {Since the stochasticity is Markovian, the flow map still corresponds to a stochastic time-dependent curve $g_t$ on the group of compositions of smooth invertible maps, i.e., the diffeomorphisms, acting on flow domain $\mathcal{D}$, see \cite{CrFlHo2017}. }

Thus, following  \cite{Kr1994}, \cite{MiRo2004}, and \cite{Ho2015}, we may begin by assuming that the stochastic paths $x_t=g_t(X) $ solve a {\it Lagrangian} stochastic differential equation (SDE) with \emph{prescribed} spatially dependent function $\xi_t=\xi(g_t(X))$
\begin{align}
dg_t(X) = u_t(g_t(X))dt + \xi(g_t(X))\circ dW (t) \,,\quad\hbox{with}\quad 
g_0(X) = X \in \mathcal{D}
\,,\label{dg-Lag}
\end{align}
where $g_t:\mathcal{D}\to\mathcal{D}$ is a spatially smooth invertible map depending on time, $t\in\mathbb{R}$. 
The corresponding {\it Eulerian} stochastic {\it velocity} decomposition is given 
in terms of \textit{cylindrical noise}, introduced in \cite{Sc1988}, as
\begin{align}
dg_tg_t^{-1}(x) = u_t(x)\,dt + \xi(x)\circ dW (t) 
\,,\quad\hbox{with}\quad 
g_0(X) = X \in \mathcal{D}
\,.
\label{dx_t-Eul}
\end{align}

In this approach, the particle-and-field duality of the Lagrangian and Eulerian descriptions of continuum fluid motion is still available, even through the fluid parcel motions are stochastic. This means the symmetry of the Eulerian description under relabelling of Lagrangian fluid parcels persists, even when the Lagrangian fluid paths are stochastic, so that the Lagrange-to-Euler fluid map is also stochastic. This relabelling symmetry implies the Kelvin circulation theorems for fluid dynamics models \cite{HoMaRa1998}. Of course, these remarks generalise to any number of dimensions.

\paragraph{What does this paper do?} 
Given the stochasticity in the flow map, how does one derive the corresponding Eulerian equations of continuum motion? The paper lays out a geometric framework for deriving stochastic Eulerian motion equations for geophysical fluid dynamics (GFD) using the method of symmetry reduction for a modified Hamilton's principle for fluids with advected quantities. 

Although the paper is based on previous developments in \cite{Kr1994}, \cite{MiRo2004}, and \cite{Ho2015}, it introduces \textit{two new approaches for incorporating non-stationary statistics due to flow dependence}, as seen in the NOAA drifter data shown in Figures 1.1, 1.2 and 1.3, and analysed in \cite{Sy-etal2016}. 

In particular, the paper allows for flow dependence of the eigenvectors $\xi_t$ of the spatial correlations in the stochastic process in \eqref{dx_t-Eul}. To obtain this flow dependence, we postulate two different approaches for allowing the eigenvectors for the stochastic process to \emph{evolve} along with the advected quantities. These two models are each modifications of the approach in \cite{Ho2015}. The two models provide alternative systematic avenues for forecasting with evolving time-dependent statistics, such as those seen in the NOAA drifter data analysed in \cite{Sy-etal2016}, rather than using spatially dependent but steady statistics. While steady statistics may be appropriate for climate science, analysis of flows at shorter time scales in forecasting weather variability, for example, may require flow-dependent, evolving statistics. Our purpose here is to present a systematic framework for modelling non-stationary statistics in stochastic fluid flows.

\paragraph{Plan of the paper.} 
In section \ref{sec-Holm2015Review}, we quickly review the approach of \cite{Ho2015} as it applies to the Euler equations of a perfect fluid. The methodology of this approach has two primary features: (1) stochastic variational principles; and (2) stochastic Hamiltonian formulations. 
These two primary features will allow us to introduce two new stochastic extensions of Geophysical Fluid Dynamics (GFD), one with advection by the drift velocity of the eigenvectors $\xi_t$ discussed in section \ref{sec_advection}, and the other with eigenvectors $\xi_t$ depending on advected fluid quantities discussed in section \ref{sec-evsadvectedquantities}. Both Stratonovich and It\^o forms of the equations are provided. We conclude by comparing the three stochastic models in terms of their Kelvin circulation theorems in section \ref{sec-conclus}.

\section{Model 1: Review of stochastic variational principles for fluids}
\label{sec-Holm2015Review}

Following the geometric approach of \cite{Ar1966}, we consider the group $G$ of volume preserving diffeomorphisms of the fluid domain $ \mathcal{D} $, as the configuration manifold for incompressible fluids. Curves $g_t \in G$ in this group describe Lagrangian trajectories $x_t=g_t(X)$ of the fluid motion. To simplify our discussion, we will take the domain $ \mathcal{D} $ to lie in $ \mathbb{R}  ^2 $ or $ \mathbb{R}  ^3 $ and neglect considerations of boundary conditions. Our developments extend easily to the case where $ \mathcal{D} $ is a manifold with smooth boundary.

The Lagrangian of the incompressible fluid is defined on the tangent bundle $TG$  of the group $G$ and is given by the kinetic energy, i.e., 
\[
L(g, v)=  \int _ \mathcal{D} \frac{1}{2}|v(X)| ^2 d ^{\,n} \!X,
\]
for $n=2,3$. By a change of variables, we note that $L$ is right invariant: $L(gh, vh)=L(g, v)$, for all $h$ in $G$. We can thus write $L(g, v)= \ell( v g ^{-1} )$, where $\ell: \mathfrak{g}  \rightarrow \mathbb{R}$. That is, we write
\begin{equation}\label{ell} 
\ell(u)= \int_ \mathcal{D}\frac{1}{2} | u(x) | ^2 d ^n x,
\end{equation} 
as the reduced Lagrangian defined on the Lie algebra $ \mathfrak{g}  $ of $G$, given by the space of divergence free vector fields, denoted $u:=v g ^{-1}\in \mathfrak{g}$.

Hamilton's principle $ \delta \int_0^TL(g, \dot g) dt=0$ yields the Euler equations in Lagrangian description, i.e., geodesics on $G$, whereas the variational principle induced on $\ell$, the Euler-Poincar\'e principle, yields the Euler equations in their standard spatial description.

In particular, the stochastic variational principle in \cite{Ho2015} is obtained by selecting a space $V$ of tensor fields on $ \mathcal{D} $, denoted $q(t,x)$, on which the group $G$ acts linearly by the pull-back operation
\begin{equation}\label{pull_back_action} 
q \in V\mapsto g ^\ast q \in V,
\end{equation} 
which is the natural transport operation of tensor fields by the fluid motion.
The associated Lie algebra action of divergence free vector fields $ u\in \mathfrak{g}  $ is given by the Lie derivative as 
\[
q \in V\mapsto  \pounds _ u q :=\left.\frac{d}{d\varepsilon}\right|_{\varepsilon=0} g_ \varepsilon ^\ast q \in V,
\]
where $ g_ \varepsilon $ is the flow of $u$. We fix a space $V ^\ast $ of tensor fields in nondegenerate duality with $V$ relative to the $L ^2$ pairing
\[
\left\langle p, q \right\rangle _V=\int_\mathcal{D}  \left\langle p(x),q(x) \right\rangle d ^n x.
\]

\begin{remark}[Determining the correlation eigenvectors $ \xi _i (x) $]\rm\label{corr-eofs}
How may the eigenvectors $ \xi _i (x) $ be found in practice? One approach for determining them would be based on running an ensemble of computer simulations of the flow started with observed initial velocity data measured in the NOAA data base at each grid point $X$ on a computational mesh erected in the flow domain. This computational mesh would be significantly coarser that the resolution of the data. An ensemble of these computer runs would produce a distribution of simulated end points $X^r=U^r\,dt$ on the coarse mesh, where $r$ labels the ensemble member. The distribution of differences $\Delta X^r$ of the simulations from the observed data around each coarse grid point would be modelled (up to some specified tolerance) as a local stochastic process given by $dX = \sum_{j=1}^N \xi_j(X)\circ dW_j(t)$, where $\{\xi_j(X)\}$ is a set of orthogonal functions whose squared amplitudes are given by $\lambda_j$ (kinetic energy) for the spectrum of the correlation tensor $C(X,Y) = \xi(\{X\}) \otimes \xi^T(\{Y\})$ where vectors $\{X\}$ and $\{Y\}$ represent all of the points on the coarse grid. Finally, the choice of tolerance determines the choice of $N$. See \cite{CoCrHoPaSh2017} for an application of this stochastic data assimilation approach, in which the ``observed data'', or ``truth'',  is given by a highly resolved numerical simulation. { In practice, the noise correlation eigenvectors must be inferred from observed data, on a case by case basis which will depend on the quality and completeness of the data. However, the inference of velocity-velocity correlation tensors is standard practice in fluid turbulence experiments, cf. \cite{Berkooz1993}.} 
\end{remark}

Given $N$ time independent divergence free vector fields $ \xi _i(x)$, $i=1,...,N$, the stochastic variational principle in \cite{Ho2015} is formally written as
\begin{equation}\label{Clebsch_VP} 
\delta\int_0^T \Big[\ell(u)dt + \left\langle p, dq + \pounds_{dx_t}q\right\rangle_V\Big]=0,
\end{equation} 
with respect to variations $ \delta u$, $ \delta q$, $ \delta p$ and where $dx_t$ is defined as
\begin{equation}\label{def_dx} 
dx_t:=u(t,x) dt+ \sum_{i=1}^N\xi_i(x)\circ dW_i(t)\,,
\end{equation} 
in which the vector fields $\xi_i(x)$ represent spatial correlations of the stochasticity and $dW_i(t)$ are independent Brownian motions, introduced in the Stratonovich sense. 
The stationarity conditions are computed by fixing a space $ \mathfrak{g}  ^\ast $ in nondegenerate duality with $ \mathfrak{g}$, relative to a pairing $ \left\langle m, u \right\rangle _ \mathfrak{g}  $, $u \in\mathfrak{g}  $, $m \in\mathfrak{g}  ^\ast $.
The variations of \eqref{Clebsch_VP} in $\delta u$, $\delta p$ and $\delta q$ yield, respectively, the conditions
\begin{equation}\label{conditions} 
\frac{\delta \ell}{\delta u}= p \diamond q,\qquad dq+ \pounds _{dx_t}q=0, \qquad dp- \pounds ^\mathsf{T}_{dx_t}p=0,
\end{equation} 
where $ p \diamond q \in \mathfrak{g}  ^\ast , \frac{\delta \ell}{\delta u} \in\mathfrak{g}  ^\ast  $, and $ \pounds ^\mathsf{T}_up\in V^\ast $ are defined as 
\begin{equation}\label{definition_diamond}
\left\langle p \diamond q, u \right\rangle _ \mathfrak{g}  = \left\langle p,\pounds _ u q \right\rangle _V  ,  \qquad \left\langle \frac{\delta \ell}{\delta u}, \delta u \right\rangle _\mathfrak{g}  := \left.\frac{d}{d\varepsilon}\right|_{\varepsilon=0} \ell(u+ \varepsilon \delta u) , \qquad \left\langle\pounds _u^\mathsf{T} p, q \right\rangle _V= \left\langle p,\pounds _u q \right\rangle _V ,
\end{equation} 
for $q \in V$, $p \in V ^\ast $, and $u, \delta u \in \mathfrak{g}  $.
The conditions \eqref{conditions} imply the following stochastic equation:
\begin{equation}\label{Stoch_EP}
d \frac{\delta \ell}{\delta u}+ \operatorname{ad}^* _{dx_t} \frac{\delta \ell}{\delta u} =0,
\end{equation}
where $\operatorname{ad}^*_u: \mathfrak{g}^\ast \rightarrow \mathfrak{g}  ^\ast $ denotes the coadjoint operator defined by $ \left\langle \operatorname{ad}^*_u m, v \right\rangle _ \mathfrak{g}  =\left\langle m, [u,v] \right\rangle_\mathfrak{g}$, with $[u,v]= v\cdot \nabla u-u\cdot \nabla v$, and $ dx_t$ is given in \eqref{def_dx}. 

The notations used in \eqref{Stoch_EP} are general enough to make this equation valid for any Lie group $G$ and Lagrangian $ \ell: \mathfrak{g}  \rightarrow \mathbb{R} $. See \cite{ArCaHo2017} for a parallel treatment of Model 1 for the rigid body and the group $SO(3)$, as well as for the heavy top, which involves advected quantities arising from symmetry breaking from $SO(3)$ to $SO(2)$. 

Upon choosing for $ \mathfrak{g}  ^\ast $ the space of divergence free vector fields on $ \mathcal{D} $, i.e., $ \mathfrak{g}  ^\ast = \mathfrak{g}  $, and the duality pairing 
\[
\left\langle m,u \right\rangle _ \mathfrak{g}  = \int_ \mathcal{D} m(x) \!\cdot\! u(x) \,d^nx\,,
\]
the coadjoint operator is $\operatorname{ad}^*_u m= \mathbb{P}  ( u\cdot \nabla m+ \nabla u^\mathsf{T} \cdot m)$, where $ \mathbb{P}$ is the Hodge projection onto divergence free vector fields. With the Lagrangian \eqref{ell}, the stochastic Euler equation \eqref{Stoch_EP} becomes, in 3D,
\begin{equation}\label{stoch_3D_Euler} 
d u+ \mathbb{P}  ( u \cdot \nabla u) dt+\sum_{i=1}^N \mathbb{P}  ( \operatorname{curl} u \times \xi_i  ) \circ dW _i (t) =0\,.
\end{equation} 
{
Equation \eqref{stoch_3D_Euler} can be written equivalently in vorticity form as
\begin{equation}\label{stoch_vort_eqn} 
d\omega+({dx_t}\cdot\nabla)\omega - (\omega\cdot\nabla){dx_t}=0,
\end{equation} 
where $\omega={\rm curl}\, u$ is the vorticity and the stochastic vector field ${dx_t}$ is given in equation \eqref{def_dx}. 
}

In 2D we identify the space of divergence free vector fields with the space of functions on $ \mathcal{D} $ modulo constants, the stream functions $ \psi $. As explained in \cite{MaWe1983}, the dual space $ \mathfrak{g}  ^\ast $ is identified with functions on $ \mathcal{D} $ with zero integrals, the absolute vorticities $ \varpi $, via the duality pairing
\[
\left\langle \varpi  , \psi \right\rangle _ \mathfrak{g}  = \int_ \mathcal{D} \varpi (x)\psi (x) d ^2 x.
\]
The absolute vorticity $\varpi$ is related to the total fluid momentum $m$ as $ \varpi = \operatorname{curl} m \cdot \mathbf{z} $. For instance, for the Euler equation, the absolute vorticity coincide with the vorticity $ \omega = \operatorname{curl} u \cdot \mathbf{z}$, whereas for the rotating Euler equation, we have $ \varpi  = \operatorname{curl} u \cdot \mathbf{z}+f= \omega +f$, where $f$ is Earth's frequency rotation.

In 2D, the stochastic Euler equation \eqref{Stoch_EP} becomes
\begin{equation}\label{stoch_2D_Euler} 
d\omega + \{ \omega , \psi \} dt+ \sum_{i=1}^ N\{ \omega , \psi_i \}\circ dW _i (t) =0,
\end{equation} 
where, for two functions $f,g$ on $\mathcal{D} $, the function $\{f,g\}$ is the Jacobian defined by $ \{f,g\}:= \partial _{x _1 } f\partial _{x_2} g - \partial _{x _2 } f\partial _{x_1} g  $, with $x=(x_1 , x _2 )$. In \eqref{stoch_2D_Euler}, $ \psi (t,x)$ is the stream function of the fluid velocity $u(t,x)$, $ \omega (t,x)=- \Delta \psi (t,x)$ is its vorticity, and the functions $ \psi _i (x)$ are the stream functions of the divergence free vector fields $ \xi _i (x)$. The deterministic Euler equations are recovered in \eqref{stoch_3D_Euler} and \eqref{stoch_2D_Euler} when $ \xi _i =0$, for all $=1,...,N$.

\medskip

We shall now make two crucial observations about this approach, that will allow us later to extend the approach further to develop the other two stochastic models treated in the paper.

\paragraph{Observation 1: Stochastic variational principles.}
Knowing that the deterministic Euler equations in the Lagrangian fluid description arise from the Hamilton principle
\begin{equation}\label{HP} 
\delta \int_0^T L(g, \dot g) dt=0\,,
\end{equation} 
for the right invariant Lagrangian $L:T G\rightarrow \mathbb{R}$ given by the kinetic energy, 
we expect the equations \eqref{Stoch_EP} to arise, in the Lagrangian description, via a stochastic extension of Hamilton's principle \eqref{HP}. This is indeed the case if one proceeds formally here and below by considering the \textit{stochastic Hamilton-Pontryagin  (SHP) principle}%
\footnote{This is a variational principle on the Pontryagin bundle $TG \oplus T^*G \rightarrow G$, defined as the vector bundle over $G$ with vector fiber at $g \in G$ given by $T_gG \oplus T^*_gG$.}
\begin{equation}\label{SVP}
\delta \int_0^T \Big[L(g, v)dt+ \big\langle \pi , dg- vdt - \sum_{i=1}^N\xi_ig\circ d W _i(t) \big\rangle \Big]=0\,,
\end{equation}
for variations $ \delta g$, $ \delta v$, $ \delta \pi $. The variables $v$ and $ \pi $ are, respectively, the material fluid velocity and material fluid momentum. In \eqref{SVP}, $ \left\langle\,\cdot \,, \,\cdot\,\right\rangle $ denotes the pairing between elements in $T_g^\ast G$, and $T_g G$, the cotangent and tangent space to $G$ at $g$. 
{
The notation $ \xi _i g$ indicates the composition of the vector field $ \xi _i $ on the right by the diffeomorphism $g$.  
}
Stochastic Hamilton-Pontryagin principles (SHP) have been considered for finite dimensions in \cite{BROw2009}. 
{The present paper considers SHP in infinite dimensions for the first time. The SHP affords a systematic derivation of the stochastic equations that preserves their deterministic mathematical properties, both geometrical and analytical.}

In the present paper, we shall consider stochastic variational principles in infinite dimensions only in a formal sense for the purpose of modelling time-dependent spatial correlations. The corresponding questions in analysis, for example, the questions of local in time existence and uniqueness of solutions answered in \cite{CrFlHo2017} for the stochastic 3D Euler fluid model, will all be left open for the two new stochastic geometric fluid models that are introduced in this paper.

Note that \eqref{SVP} imposes the stochastic process \eqref{dg-Lag} as a constraint on the variations by using the Lagrange multiplier $ \pi $. From the $G$-invariance of both the Lagrangian and the constraint, this principle can be equivalently written formally in the reduced Eulerian description as
\begin{equation}\label{Reduced_SVP}
\delta \int_0^T \Big[\ell(u)dt
+ \big\langle m, dgg^{-1}- udt 
- \sum_{i=1}^N\xi_i\circ d W _i(t) \big\rangle _ \mathfrak{g}  \Big]=0\,,
\end{equation}
with respect to variations $ \delta u, \delta g, \delta m $, and where $u= v g ^{-1} \in \mathfrak{g}   $, $ m = \pi g ^{-1} \in \mathfrak{g}  ^\ast  $.
This is the \textit{reduced stochastic Hamilton-Pontryagin (RSHP) principle}.

One then directly checks that the stochastic variational principle \eqref{Reduced_SVP} also yields the stochastic equation \eqref{Stoch_EP}. Thus, the two variational principles \eqref{Clebsch_VP} and \eqref{Reduced_SVP} both yield the same stochastic equations. Moreover, in absence of stochasticity, equation \eqref{SVP} recovers the Hamilton-Pontryagin principle for Lagrangian mechanics, see \cite{YoMa2006}.

\medskip

\begin{remark}{\rm
The RSHP principle in \eqref{Reduced_SVP} has several interesting properties: (i) it allows a formulation of reduction by symmetry in the stochastic context; (ii) it  does not need the introduction of the extra advected quantities $q,p$; and (iii) it does not restrict the values of the Eulerian fluid momentum $m \in \mathfrak{g}  ^\ast $ to be of the form, $m=p\,\diamond \,q$. In addition, as we will show later, the unreduced SHP principle  \eqref{SVP} allows us to consistently implement the new Model 2 and Model 3, in which the spatial correlation eigenvectors $ \xi_i(x)$ which are fixed functions of the spatial coordinates in Model 1 become time dependent through their flow dependence in Model 2 and Model 3.}
\end{remark}

\paragraph{Observation 2: Stochastic Hamiltonian formulations.} We note that the SHP principle \eqref{SVP} can be equivalently written as 
\begin{equation}\label{Ham_SVP_1}
\delta \int _0 ^T  \Big[L(g, v)+ \big\langle  \pi , dg- vdt \big\rangle- \sum_{i=1}^NH_i(g,\pi; \xi _i  )\circ d W _i(t) \Big] =0\,,
\end{equation}
for the $G$-invariant functions $H_i(\_\,,\_\,; \xi _i ):T^*G \rightarrow \mathbb{R}$ defined by
\begin{equation}\label{H_i}
H_i(g, \pi ; \xi _i ):= \langle\pi , \xi_i g \rangle= \left\langle \pi g ^{-1}  ,\xi_i\right\rangle_ \mathfrak{g}  , \;\; i=1,...,N .
\end{equation} 
The principle in \eqref{Ham_SVP_1} yields the following stochastic extension of the Euler-Lagrange equations with Lagrangian $L$:
\begin{equation}\label{Lag_form} 
d \frac{\partial L}{\partial v}=\frac{\partial L}{\partial g}dt- \sum_{i=1}^N\frac{\partial H _i }{\partial g} \circ dW _i (t) =0, \quad  dg=v dt+\frac{\partial H_i }{\partial \pi }\circ dW _i (t) , \quad \pi = \frac{\partial L}{\partial v}\,.
\end{equation} 
This is the Lagrangian description of the stochastic equations \eqref{Stoch_EP}.
Denoting by $H: T^*G \rightarrow \mathbb{R}  $,  the Hamiltonian associated to $L$ by the Legendre transform, we can rewrite these equations in stochastic Hamiltonian form
\begin{equation}\label{Ham_form} 
dg= \frac{\partial H}{\partial \pi } dt+  \sum_{i=1}^N \frac{\partial H _i }{\partial\pi }\circ d W _i (t)  , \quad d\pi = -   \frac{\partial H}{\partial g } dt- \sum_{i=1}^N  \frac{\partial H _i }{ \partial g }\circ d W _i (t)\,.
\end{equation} 
Consequently, we can call the functions $H _i $ the \textit{stochastic Hamiltonians}. Stochastic Hamiltonian systems of the form \eqref{Ham_form} have been developed in \cite{Bi1982}. The intrinsic form of equations \eqref{Lag_form} and \eqref{Ham_form} on the Lie group $G$ would require introducing a covariant derivative. The formulation above is only valid locally.

These equations can be written in terms of the \textit{canonical Poisson bracket} $\{\,\cdot\,,\,\cdot\,\}_{\rm can}$ on $T^*G$ as
\begin{equation}\label{Poisson_form} 
dF = \{F,H\} _{\rm can}dt+\sum_{i=1}^N \{F, H_i \}_{\rm can}\circ d W _i  (t) \,,
\end{equation} 
for arbitrary functionals $F=F(g, \pi ): T^*G \rightarrow \mathbb{R}  $.

Consistently with this observation, we note that the stochastic equation \eqref{Stoch_EP} can also be written in Hamiltonian form as
\begin{equation}\label{LP_stoch}
d m + \operatorname{ad}^*_{ \frac{\delta h}{\delta m }}m \,dt+ \sum_{i=1}^N \operatorname{ad}^*_{ \frac{\delta h _i }{\delta m }  } m \circ dW _i (t) =0\,,
\end{equation} 
where $h: \mathfrak{g}  ^\ast \rightarrow \mathbb{R}  $ and $h _i :\mathfrak{g}  ^\ast \rightarrow \mathbb{R}$ are the reduced Hamiltonians associated to $H$ and $H _i $ in \eqref{H_i}, i.e., $H(g, \pi )= h( \pi g ^{-1} )$ and $H _i (g,\pi ;\xi _i )= h_i (\pi  g^{-1} )$. We find
\begin{equation}\label{ed_hi}
h(m)= \int_ \mathcal{D} \frac{1}{2} |m(x)| ^2 d^nx\quad\text{and}\quad  h _i ( m )=  \left\langle m , \xi_i\right\rangle_ \mathfrak{g} =\int_ \mathcal{D} m(x)\!\cdot \! \xi _i (x) \,d^nx.
\end{equation}
The expression \eqref{LP_stoch} is the reduced (or Eulerian) formulation of the Hamiltonian formulation \eqref{Ham_form}. 

In terms of the Lie-Poisson bracket $\{\,,\}_{\rm LP}$ on $ \mathfrak{g}  ^\ast $, given by
\[
\{f,h\}_{\rm LP}( m )= \left\langle m , \left[\frac{\delta f}{\delta m }, \frac{\delta h}{\delta m }\right] \right\rangle_ \mathfrak{g}  ,
\]
equation \eqref{Stoch_EP} and hence \eqref{LP_stoch} can be formulated in the Stratonovich-Lie-Poisson form
\begin{equation}\label{LP_form} 
df = \{f,h\} _{\rm LP}dt+\sum_{i=1}^N \{f, h_i \}_{\rm LP}\circ d W _i(t)  \,,
\end{equation}
for arbitrary functions $f: \mathfrak{g}  ^\ast \rightarrow \mathbb{R}  $,
which is just the reduced form of \eqref{Poisson_form}.

For example, the stochastic 2D Euler equations \eqref{stoch_2D_Euler}, can be written in the Stratonovich-Lie-Poisson form \eqref{LP_form} with  the Lie-Poisson bracket written on the space of vorticities as \cite{MaWe1983}
\[
\{f,g\}_{\rm LP}( \omega )= \int_ \mathcal{D}  \omega \left\{ \frac{\delta f}{\delta \omega },\frac{\delta g}{\delta \omega }   \right\} d^2 x
\]
and with the stochastic Hamiltonians
\[
h _i ( \omega )=\int _ \mathcal{D} \omega (x) \psi _i (x) \, d ^2 x.
\]

\paragraph{Aim of the paper.}
By exploiting the two observations discussed above, this paper will develop two new stochastic models by appropriate modifications of the stochastic Hamiltonians $H _i $ and of their symmetries.

\section{Model 2: Frozen-in correlations for non-stationary statistics}\label{sec_advection}

As discussed in the Introduction, we consider a model in which the eigenvectors $ \xi _i (X)$ are advected by the flow map $g_t$, giving the time dependent vector fields $ \zeta _i  (t,x)$. Mathematically, the advection of a vector field by a smooth invertible map $g_t$  corresponds to the \textit{push-forward operation},
\begin{equation}\label{PF} 
\zeta_i (t)=  (g_t) _ {\ast}\xi _i 
\,,
\end{equation} 
meaning that
\[
\zeta _i( t,g_t(X))= Dg_t (X) \cdot \xi _i ( X), \quad \text{for all $ X\in \mathcal{D} $} 
\,.
\]
From a Lie group point of view, the push-forward is given by the adjoint action Ad$:G\times \mathfrak{g}\to  \mathfrak{g}$ of the Lie group $G$ on its Lie algebra $ \mathfrak{g}$, that is, 
\[
\zeta _i(t) = (g _t)_\ast \xi _i = \operatorname{Ad}_{g_t} \xi _i \,. 
\]
The notation used in this section is general enough to make our developments valid for any Lie group $G$. In our examples, we will take $G$ to be the either the group of diffeomorphisms or the group of volume preserving diffeomorphisms. The corresponding coadjoint operators are
\begin{equation}\label{formula_adstar} 
\operatorname{ad}^*_u m=  u\cdot \nabla m+ \nabla u^\mathsf{T} \cdot m+ m \operatorname{div}u\qquad\text{and}\qquad  \operatorname{ad}^*_u m= \mathbb{P}  ( u\cdot \nabla m+ \nabla u^\mathsf{T} \cdot m).
\end{equation}

The stochastic model considered here for advection of the eigenvectors is obtained by modifying the expression and the symmetries of the stochastic Hamiltonians $H_i(g,\pi ;  \xi _i )$ in \eqref{Ham_SVP_1}. Namely, given $N$ vector fields $ \xi _i $, $i=1,...,N$, we consider the principle
\begin{equation}\label{Ham_SVP_3}
\delta \int_0^T \Big[L(g, v)dt+ \big\langle  \pi , dg- vdt \big\rangle- \sum_{i=1}^NH_i(g,\pi ;  \xi _i )\circ d W _i(t) \Big] =0,
\end{equation}
where each of the stochastic Hamiltonians $H _i (\_\,,\_\,;\xi _i ):T^*G \rightarrow \mathbb{R}$ is right invariant only under the action of the \textit{isotropy subgroup} of the eigenvector $ \xi _i $ with respect to the adjoint action (i.e., the push-forward action), namely,
\[
G_{ \xi _i }=\{ g \in G \mid \operatorname{Ad}_g \xi _i = \xi_i  \}=\{ g \in G \mid g _\ast \xi _i = \xi_i  \} \subset G\,.
\]
That is, we have $H _i (gh, \pi h;\xi _i )= H _i( g, \pi ; \xi _i )$, for all $ h \in G_{ \xi _i }$.
The SHP principle \eqref{Ham_SVP_3} yields equations in the same general form as \eqref{Ham_form}, but with stochastic Hamiltonians which are not $G$-invariant. As we will show, this difference due to symmetry breaking from $G$ to $G_{\xi_i}$ induces significant changes in the reduced Eulerian representation. 

Physically, the symmetry breaking means that the initial conditions for the correlation eigenvectors are ``frozen'' into the subsequent flow, as a property carried along with individual Lagrangian fluid parcels, and which is not exchanged with other fluid parcels.

Being only $G_{ \xi _i }$-invariant, the stochastic Hamiltonian function $H _i $ induces, in the Eulerian description, the reduced stochastic Hamiltonians
\begin{equation}\label{reduced_hi} 
h _i = h _i (m , \zeta _i ):(T^*G)/G_{ \xi _i }\simeq \mathfrak{g}  ^\ast \times \mathcal{O} _{ \xi_i } \rightarrow \mathbb{R}\,,
\end{equation} 
defined by
\begin{equation}\label{reduced_hi_def}
h _i ( m   , \zeta_i )=H _i ( g, \pi ; \xi _i ), \quad\text{for}\quad   m = \pi g ^{-1}, \;\;\zeta_i = \operatorname{Ad}_g\xi _i
\,,\end{equation} 
where $ \mathcal{O} _{\xi _i }:= \{\operatorname{Ad}_g \xi_i \mid g \in G\}=\{g _\ast  \xi_i \mid g \in G\} \subset \mathfrak{g}$ is the adjoint orbit of $ \xi _i $. The SHP principle \eqref{Ham_SVP_3} can thus be written in the reduced Eulerian form as
\begin{equation}\label{red_SVP_3}
\delta \int _0 ^T \Big[\ell(u) dt+ \big\langle m  , dg g^{-1} - u dt \big\rangle_ \mathfrak{g}  - \sum_{i=1}^Nh_i( m ,   \operatorname{Ad}_g \xi _i )\circ d W _i(t) \Big] =0\,.
\end{equation}
The stationarity conditions with respect to the variations $ \delta u$, $ \delta m$, $ \delta g$, yield\begin{align}\label{model1}  
\begin{split}
\delta u:&\quad \frac{\delta \ell}{\delta u} - m = 0
\,,\qquad
\delta m:\quad
dg g ^{-1} - u dt - \sum_{i=1}^N\frac{\delta h _i }{\delta m } \circ dW _i (t) = 0
\,,
\\ \delta g:&\quad
d \frac{\delta \ell}{\delta u}
+ \operatorname{ad}^*_ u \frac{\delta \ell}{\delta u} dt
- \sum_{i=1}^N \operatorname{ad}^*_ { \zeta _i } \frac{\delta h _i }{\delta \zeta _i } \circ dW _i (t) =0\,,
\end{split}
\end{align} 
where the advected eigenvector $ \zeta _i := \operatorname{Ad}_g \xi _i $ obeys the auxiliary equation $d\zeta_i = [dgg^{-1},\zeta_i ]$, obtained from its definition. The expression of the coadjoint operator $ \operatorname{ad} ^\ast $  is given in \eqref{formula_adstar}. Upon using the second equation in \eqref{model1}, this auxiliary equation becomes
\begin{align}\label{model1-auxeqn}  
d \zeta _i +[ \zeta _i ,u]dt+\sum_{j=1}^N\Big[ \zeta _i , \frac{\delta h _j }{\delta m }\Big] \circ dW _j (t)  =0\,. 
\end{align} 

\begin{remark}\label{remark3_1}\rm
In our applications of this model, we will always assume that the stochastic Hamiltonians $h _i $ in \eqref{model1-auxeqn} do not depend on $ m $. That is, $\delta h _j /\delta m =0$, so that the  eigenvectors $\zeta _i $ are advected only by the drift velocity, $u$. That is,
\begin{equation}\label{zeta_equation} 
d \zeta _i +[ \zeta _i ,u]dt=0\,.
\end{equation} 
This model using correlation eigenvectors frozen into the drift velocity is quite different from the model in \cite{Ho2015}, which chose $h_i(m)=\langle m\,,\,\xi_i\rangle_ \mathfrak{g} $ and all fluid properties were advected by a velocity vector field comprising the sum of both the drift component and the stochastic component. 
\end{remark}

\paragraph{Hamiltonian structure.} Denoting by $h: \mathfrak{g}  ^\ast \rightarrow \mathbb{R}  $ the Hamiltonian associated to $\ell$, the above system can be equivalently written as
\begin{equation}\label{syst_h}
\left\{
\begin{array}{l}
\displaystyle d  m + \operatorname{ad}^*_ {\frac{\delta h}{\delta m }} m \, dt- \sum_{i=1}^N \operatorname{ad}^*_ { \zeta _i } \frac{\delta h _i }{\delta \zeta _i } \circ dW _i (t) =0
\,,\\
\displaystyle d \zeta _i +\Big[  \zeta _i , \frac{\delta h}{\delta m} \Big] dt+\sum_{j=1}^N\Big[ \zeta _i , \frac{\delta h _j }{\delta m }\Big] \circ dW _j (t) =0
\,.
\end{array}
\right.
\end{equation}
One may check that this system is Hamiltonian with respect to the Poisson bracket
\begin{equation}\label{PB_S3} 
\{f,g\}_{\rm red}( m , \zeta _1 , ..., \zeta _N )=  \left\langle m , \left[\frac{\delta f}{\delta m }, \frac{\delta g}{\delta m }\right] \right\rangle_ \mathfrak{g}   + \sum_{i=1}^N \left\langle \left[ \zeta _i , \frac{\delta f}{\delta m } \right] , \frac{\delta g}{\delta \zeta _i }  \right\rangle _\mathfrak{g}  - \sum_{i=1}^N \left\langle \left[ \zeta _i , \frac{\delta g}{\delta m } \right] , \frac{\delta f}{\delta \zeta _i }  \right\rangle_\mathfrak{g}  
\end{equation} 
on $ \mathfrak{g}  ^\ast \times  \mathcal{O} _{ \xi  }\ni ( m , \zeta _1 , ..., \zeta _N )$, where $\mathcal{O} _{ \xi  } \subset \mathfrak{g}  ^N$ is the orbit of $ \xi  := ( \xi _1 , ..., \xi _N )$ under the adjoint action of $G$. This reduced Poisson bracket is inherited from Poisson reduction of the canonical Poisson bracket $\{\, \cdot\,,\,\cdot\, \}_{\rm can}$ on $T^*G$, by the isotropy subgroup $G_ \xi  \subset G$ of $\xi :=(  \xi _1 , ...., \xi _N)$. Namely, the map
\[
(g, \pi ) \in T^*G \mapsto ( m, \zeta _1,..., \zeta_N )= ( \pi g ^{-1} ,\operatorname{Ad}_g \xi _1 ,...,\operatorname{Ad}_ g \xi_N    )\in  (T^* G)/G_ \xi  \simeq \mathfrak{g}  ^\ast \times  \mathcal{O}_{ \xi  }
\]
is Poisson with respect to the Poisson brackets $\{\, \cdot\,,\,\cdot\, \}_{\rm can}$ and $\{\, \cdot\,,\,\cdot\, \}_{\rm red}$, see the Appendix. The system \eqref{syst_h} admits the Stratonovich-Poisson formulation,
\begin{equation}\label{Poisson_Strat} 
df = \{f,h\} _{\rm red}dt+\sum_{i=1}^N \{f, h_i \}_{\rm red}\circ d W _i (t)  \, ,
\end{equation} 
for arbitrary functions $f: \mathfrak{g}  ^\ast\times \mathcal{O} \rightarrow \mathbb{R}  $. Note that in \eqref{Poisson_Strat} the Hamiltonians $h$ and $h _i $, $i=1,...,N$, depend a priori on all the variables $(m ,\zeta _1 , ...,\zeta _N)$. The system \eqref{syst_h} is recovered when $h$ depends only on $ m $,  while the $ h _i $ depend only on $ m $ and $\zeta_i $ (not on $ \zeta _j $, for $j\neq i$).
The Poisson tensor at $(m, \zeta _1 ,..., \zeta _N )$ reads
\begin{equation}\label{matrix}
\left[
\begin{array}{cccc}
\displaystyle - \operatorname{ad}^*_{\square}  m& \operatorname{ad}^*_{ \zeta _1 }  & ... &  \operatorname{ad}^*_{ \zeta _N }  \\
\displaystyle - \operatorname{ad}_{ \zeta _1 }  & & &\\
\displaystyle \vdots & & 0&\\
\displaystyle - \operatorname{ad}_{ \zeta _N }  & & &
\end{array}
\right].
\end{equation}
\medskip

\begin{remark}[It\^o form]\label{Ito-remark}\rm
In the special case that ${\delta h_i}/{\delta m }=0$, as assumed in Remark \ref{remark3_1}, the It\^o form of equations \eqref{syst_h} does not introduce any additional drift terms. That is, in this special case, the It\^o form of \eqref{syst_h} is obtained by simply removing the Stratonovich symbol $(\,\circ\,)$. 
However, in the general case, if ${\delta h_i}/{\delta m }\ne0$, the It\^o form does contain additional drift terms. These additional drift terms in the It\^o form for the general case can be computed in a standard way, but the equations then may take a more complicated form. By making use of the Poisson formulation in terms of the bracket $\{\, \cdot\,,\,\cdot\, \}_{\rm red}$ in  \eqref{Poisson_Strat}, we can write the additional drift terms in the It\^o form for the general case in a concise way as
\begin{equation}\label{Ito_Poisson} 
df =\left(  \{f,h\} _{\rm red}- \frac{1}{2} \{h_i,\{ h_i,f\}_{\rm red}\} _{\rm red} \right) dt+\sum_{i=1}^N \{f, h_i \}_{\rm red}\circ d W _i (t)  \,.\end{equation} 
\end{remark}

\begin{example} [Incompressible 2D models]\rm In the 2D incompressible case, we can identify the Lie algebra $ \mathfrak{g}  $ with the space of differentiable  functions on $ \mathcal{D} $, modulo constants. These are the stream functions, denoted $ \psi $.
We use the $L ^2 $ duality pairing $ \left\langle \varpi , \psi \right\rangle _ \mathfrak{g}  =\int_ \mathcal{D} \varpi (x) \psi (x) d^2x$ and identify $\mathfrak{g} ^\ast $ with the space of functions on $ \mathcal{D} $ with zero integral. These are the absolute vorticities, denoted $ \varpi $, as explained in \cite{MaWe1983}.

Let $ \psi _i^0(X)$ be the stream function associated to the eigenvector $ \xi _i (X)$, $i=1,...,N$. The stream function $ \psi _i (t,x)$ of the advected eigenvector $ \zeta _i (t)= (g_t) _\ast \xi _i $ is found to be
\[
\psi _i (t,g_t(X))= \psi _i ^0(X).
\]
The stochastic model \eqref{syst_h} applied to 2D incompressible fluid dynamics with Hamiltonian $h( \varpi  )$ and stochastic Hamiltonians $ h _i ( \varpi  , \psi_i )= h _i ( \psi _i  )$ is given by%
\begin{equation}\label{2D_model1} 
d\varpi+ \left\{ \varpi, \frac{\delta h}{\delta \varpi} \right\} dt+\sum_{i=1}^N\left\{ \psi_i  ,\frac{\delta h _i }{\delta \psi_i } \right\} \circ dW _i (t) =0
\,, \qquad 
d \psi _i + \left\{ \psi _i , \frac{\delta h}{\delta \varpi} \right\} dt=0,
\end{equation} 
where we have used the formula $ \operatorname{ad}^*_ \psi  \varpi =\{\varpi, \psi \}$ for the coadjoint operator for 2D incompressible fluids. For example, with the appropriate choice of the Hamiltonian,
we can write the stochastic model \eqref{2D_model1} for the following cases:
\begin{itemize}
\item[\rm (a)] 2D perfect fluid: $ \psi =\frac{\delta h}{\delta \varpi}= - \Delta ^{-1} \varpi$;
\item[\rm (b)] 2D rotating perfect fluid: $ \psi =\frac{\delta h}{\delta \varpi}= - \Delta ^{-1} (\varpi-f)$;
\item[\rm (c)] 2D rotating quasigeostrophy (QG): $ \psi =\frac{\delta h}{\delta \varpi}= -( \Delta- \mathcal{F} ) ^{-1} (\varpi-f)$;
\end{itemize}
where $ \mathcal{F} $ and $f$ denote, respectively, the square of the inverse Rossby radius and the rotation
frequency. For the stochastic Hamiltonians $h _i $, $i=1,...,N$, one may choose
\[
h_i ( \psi_i )= - \frac{1}{2} \int _ \mathcal{D} \Delta \psi _i (x)  \psi _i (x) \,d ^2 x,\;\; \text{(no sum)},
\]
in which case $ \frac{\delta h _i }{\delta \psi_i } = - \Delta\psi _i $.

Similarly to the discussion in Remark \ref{Ito-remark}, the It\^o form of equations  \eqref{2D_model1} takes the same expression, since we have chosen stochastic Hamiltonians for which $\delta h _i / \delta \varpi = 0$. In this case, one may obtain the It\^o forms by simply replacing the Stratonovich noise $ \circ \, dW _i (t)$ by the Ito noise $ dW _i (t) $ without modifying the drift terms.
\end{example}

\begin{remark}[Conserved correlation enstrophies]\rm
The stochastic model \eqref{2D_model1} does not preserve the well-known vorticity enstrophies $\Lambda ( \omega)=\int _ \mathcal{D} \Phi ( \omega) d^2 x\,,$  which are preserved in the deterministic case. 
However, the stochastic equation for $\psi_i$ in \eqref{2D_model1} does preserve the following \textit{correlation enstrophy functionals}
\[
\Lambda _i ( \psi _i )=\int_ \mathcal{D}  \Phi ( \psi_i ) d^2 x\,,
\]
when the stochastic model \eqref{syst_h} is applied to 2D incompressible fluid dynamics.
In general, for the case $h=h(m)$ and $ h _i = h _i ( \zeta _i )$, by equation \eqref{syst_h}, the corresponding functionals $  \Lambda_i ( \zeta _i )$ verify
\[
d \Lambda _i = \int_ \mathcal{D}  \{ \Lambda _i ,h\}_{\rm red}d^2x\,dt 
= - \int _ \mathcal{D} \left\langle \left[  \zeta _i , \frac{\delta h}{\delta m} \right] ,\frac{\delta  \Lambda_i }{\delta \zeta _i } \right\rangle _ \mathfrak{g}d^2x\,  dt 
\]
which vanishes for 2D incompressible fluids after integration by parts and imposition of homogeneous boundary conditions. 
\end{remark}

\medskip

\begin{example} [3D incompressible Euler]\rm We consider the stochastic Hamitonians
\begin{equation}\label{h_i_3D} 
h _i (\zeta _i )= \int _ \mathcal{D} F_i ( \zeta _i (x)) d ^3 x, \quad i=1,...,N,
\end{equation} 
where $F_i$ are smooth functions. Upon using the formula from \eqref{formula_adstar} for incompressible flows,
\begin{equation}\label{ad-star_3D} 
 \operatorname{ad}^*_u m= \mathbb{P}  ( u \cdot \nabla m+ \nabla u^\mathsf{T} \cdot m)= \mathbb{P}  ( \operatorname{curl}m\times  u) 
 \,,
 \end{equation}  
where $ \mathbb{P}  $ is the Hodge projector onto divergence free vector fields, the stochastic model \eqref{syst_h} reads
\begin{equation}\label{3D_Euler}
\left\{
\begin{array}{l}
\vspace{0.2cm}\displaystyle du+ \mathbb{P}  ( u\cdot \nabla u ) dt=\sum_{i=1}^N\mathbb{P}\left(  \operatorname{curl} \frac{\delta F _i}{\delta \zeta_i }\times  \zeta _i  \right)\circ dW _i (t) \\
d\zeta _i + \operatorname{curl}( \zeta_i \times u)dt=0.
\end{array}
\right.
\end{equation} 
The stochastic terms can be written equivalently with the help of the stress tensors
\begin{equation}\label{stoch_stress} 
\sigma_i=  \zeta_i \otimes \frac{\partial F _i }{\partial \zeta_i }+ F  _i ( \zeta _i )\mathbf{I} , \quad  \text{i.e.,} \quad (\sigma _i)_b^a = \zeta_i ^a  \otimes \frac{\partial F _i }{\partial \zeta ^b _i }- F _i ( \zeta ) \delta ^a _b
\,,
\end{equation} 
and pressures $ p _i $, $i=1,...,N$. 
{With these definitions, equation \eqref{3D_Euler} becomes 
\begin{equation}\label{stoch_motion} 
du+ (u\cdot \nabla u+ \nabla p)dt= \sum_{i=1}^N(\operatorname{div} \sigma _i +\nabla p _i ) \circ     dW _i (t),
\end{equation} 
where the divergence is defined as $(\operatorname{div} \sigma_i ) _b = \partial_a ({\sigma_i}) ^a _b $ for all $i$, and the individual $p_i$ are each found by solving a Poisson equation, with boundary conditions given by $\hat{n}_a (\operatorname{div}{\sigma_i}) ^a = 0$. } Recall that the vector fields $\zeta _i (t, x)$ are obtained from the given eigenvectors field $ \xi _i (X)$ by the push-forward operation \eqref{PF}.
As mentioned earlier in Remark \ref{Ito-remark}, the It\^o forms of the equations  have the same expression.
\medskip

\begin{remark}\label{remark} \rm
The three dimensional system \eqref{3D_Euler} is reminiscent of incompressible magnetohydrodynamics (MHD), except it has a stochastic ``$J\times B\,$'' force depending on the sum over \textit{all} of the $\zeta_i$. In following this analogy with MHD, we may introduce vector potentials $ \alpha _i$ by writing $\zeta _i=:{\rm curl\,} \alpha _i$. Having done so, one notices that evolution under the stochastic system  \eqref{3D_Euler} preserves the integrals
\[
\Lambda_i 
= \int_{\mathcal{D}} \alpha _i\cdot \zeta _i \,d^3x
\quad\hbox{(No sum).}
\]
\begin{proof}
By the second equation in the equation set \eqref{3D_Euler}, we have
\[
d\Lambda_i = d \int_{\mathcal{D}} \alpha _i\cdot \zeta _i \,d^3x 
= -2\int_{\mathcal{D}} \alpha_i \cdot 
\operatorname{curl}( \zeta_i \times u)\,d^3x \,dt= 0
\,.\]
\end{proof}
The $\Lambda_i$ integrals are topological quantities known as \emph{correlation helicities} which measure the number of linkages of the lines of each vector field $\zeta_i$ with itself.  Conservation of the correlation helicity $\Lambda_i$ means that evolution by the stochastic system  \eqref{3D_Euler} cannot unlink the linkages of each divergence free vector field $\zeta_i$ with itself. This conclusion is the analogue of conservation of magnetic helicity in MHD. 
\end{remark}

\end{example}

\paragraph{Inclusion of additional advected tensor fields.} More generally, suppose that the fluid model involves a tensor field $q(t,x)$ advected by the fluid flow as in \eqref{pull_back_action}. The evolution of this advected field is this given by $q(t)= (g_t)_\ast q_0 $, where $q_0(X)$ is the initial value and $g_t \in G$ is the fluid flow. In this case, the variational principle is written in reduced Eulerian form as
\begin{equation}\label{red_SVP_3_advected}
\delta \int _0^T\Big[\ell(u, g_\ast q_0) dt+ \big\langle m  , dg g^{-1} - u dt \big\rangle_ \mathfrak{g}  - \sum_{i=1}^Nh_i(m ,   \operatorname{Ad}_g \xi _i )\circ d W _i(t) \Big] =0,
\end{equation}
for variations $ \delta u$, $ \delta m$, and $ \delta g$. The stationarity conditions yield the same first two equations of  \eqref{model1}, whereas the third one becomes
\begin{equation}\label{red_SVP_3_advect-relat}
d \frac{\delta \ell}{\delta u}+ \operatorname{ad}^*_ u \frac{\delta \ell}{\delta u} dt- \sum_{i=1}^N \operatorname{ad}^*_ { \zeta _i } \frac{\delta h _i }{\delta \zeta _i } \circ dW _i (t) = \frac{\delta \ell}{\delta q}  \diamond q \,dt.
\end{equation}
From its definition, the quantity $q (t)= (g_t)_\ast q_0$ verifies $ dq+ \pounds _{dg g ^{-1} } q=0$. For the case that $\delta h _i /\delta m=0$, this quantity is governed by the ordinary advection equation,
\begin{equation}\label{red_SVP_3_advect-eqn}
dq+ \pounds_u q \, dt=0.
\end{equation}
\begin{remark}[It\^o form]\rm
For the case that $\delta h _i /\delta m=0$, passing to the Ito formulation does not introduce any change in the drift terms.
\end{remark}

\begin{example} [Rotating shallow water]\rm 
Equations \eqref{red_SVP_3_advect-relat} and  \eqref{red_SVP_3_advect-eqn} for the inclusion of such advected quantities into Model 2 may be illustrated with the example of the rotating shallow water equation.
In this case $ \mathcal{D} $ is a two dimensional domain and $G$ is the group of diffeomorphisms of $ \mathcal{D} $. Let us denote by $\eta(t,x)$ the water depth, by $B(x)$ the bottom topography, by $R(x)$ the Coriolis vector field, and $x=(x _1 , x _2 ) \in \mathcal{D} $. The Lagrangian of the rotating shallow water system is
\begin{equation}\label{l_RSW} 
\ell(u,\eta)= \int_ \mathcal{D}  \Big[\frac{1}{2} \eta|u| ^2 + \eta R \cdot u - \frac{1}{2} g(\eta-B) ^2 \Big] d ^2 x,
\end{equation} 
where $g$ is the gravity acceleration. Taking the stochastic Hamiltonians $h _i ( \zeta _i )$ in \eqref{h_i_3D}, the stochastic variational principle in \eqref{red_SVP_3_advected} produces the equations
\begin{equation}\label{RSW}
\left\{
\begin{array}{l}
\displaystyle du+ ( u\cdot \nabla u+ \operatorname{curl}R \times u + g \nabla (\eta-B)) dt=  \frac{1}{\eta}\sum_{i=1}^N\operatorname{div} \sigma _i \circ dW _i (t)\\
\vspace{0.2cm}\displaystyle d\eta + \operatorname{div}(\eta u)dt=0, \qquad  d\zeta _i +[ \zeta_i ,u]dt=0,
\end{array}
\right.
\end{equation}
where the stochastic stress $ \sigma _i $ is defined as above in \eqref{stoch_stress}. The effect of $ \sigma _i $ can be seen by writing the Kelvin circulation theorem obtained by integrating the first equation in 
\eqref{RSW} around a loop $c(u)$ moving with the drift velocity $u(t,x)$, to find
\[
d\oint_{c(u)} \big(u+R\big)\cdot dx = \sum_{i=1}^N\oint_{c(u)} \Big(
\frac{1}{\eta}\operatorname{div} \sigma _i  \circ dW _i (t)  
\Big) \cdot dx.
\]
Thus, the total stochastic stress generates circulation of the total velocity $(u+R)$ around any loop moving with the relative fluid velocity $u$ in the rotating frame.
\end{example}
 
\begin{example}[Rotating compressible barotropic fluid]\rm 
The above developments easily extend to other fluid models such as  compressible barotropic fluid flow in a rotating frame, whose Lagrangian is 
\begin{equation}\label{Lagr_comp} 
\ell(u, \rho ) = \int_ \mathcal{D}\rho \Big( \frac{1}{2} | u| ^2 + u \cdot R - e(\rho ) - gz \Big) d ^3 x,
\end{equation} 
where $z=x_3$ is the vertical coordinate, $ \rho $ is the mass density and $e$ is the specific internal energy. With the choice \eqref{h_i_3D}, one gets from \eqref{red_SVP_3_advect-relat}, the stochastic balance of momentum
\[ 
du+ \Big( u \cdot \nabla u+ \operatorname{curl} R \times u +   g \mathbf{z} + \frac{1}{ \rho }\nabla p  \Big) dt = \frac{1}{ \rho }\sum_{i=1}^N\operatorname{div} \sigma _i \circ dW _i (t),
\]
with advection equations 
\[
d \rho + \operatorname{div}( \rho u) dt=0
\quad\hbox{and}\quad
d\zeta _i +[\zeta_i ,u]dt=0
\,.\]
\end{example}

\section{Model 3: Eigenvectors depending on advected quantities}
\label{sec-evsadvectedquantities}

The stochastic model detailed in this section applies to fluids with advected quantities. As for the model described in \S\ref{sec_advection}, the model in this section also introduces a modification of the symmetries of the stochastic Hamiltonians $H _i $ in the SHP principle \eqref{Ham_SVP_1}.
Namely, in the presence of an advected tensor field $q(t,x)$, evolving as by the push-forward $q(t)= (g_t) _\ast q _0$, we assume that $H _i $ depends on $ q _0 $. This dependence breaks the $G$-symmetry of $H _i $. In particular, we shall take $H _i (\_\,,\_\,; q_0 ):T^*G \rightarrow \mathbb{R}  $ to be a $G_{q _0 }$-invariant function, i.e., 
\[
H _i (gh, \pi h;  q_0 )= H _i (g, \pi ;  q_0 ), \quad \text{for all} \quad h \in G_{q _0 },
\]
where $G_{q _0 }=\{ g \in G\mid g^\ast q_0  = q_0 \} \subset G$ is the isotropy subgroup of the tensor field $  q_0 $ under the pull-back action.
This is similar to the symmetry assumed on the Lagrangian $L(\_\,,\_\,; q _0 ): TG \rightarrow \mathbb{R}  $ for such fluids, studied in \cite{HoMaRa1998}.
With this assumption, the SHP principle \eqref{Ham_SVP_1} becomes
\begin{equation}\label{Ham_SVP_2}
\delta \int_0^T \Big[L(g, v; q _0 )+ \big\langle  \pi , dg- vdt \big\rangle- \sum_{i=1}^NH_i(g,\pi; q _0 )\circ d W _i(t) \Big] =0\,.
\end{equation}
If we also assume that $H _i $ is linear in the material fluid momentum $ \pi $, then it necessarily takes the form
\begin{equation}\label{H_i_2} 
H _i ( g, \pi ; q _0 )= \left\langle \pi  g^{-1} , \xi _i ( g _\ast q _0 ) \right\rangle_ \mathfrak{g}  
\end{equation} 
for a function $ \xi_i : \mathcal{O} _{ q _0 } \rightarrow \mathfrak{g}$ defined on the $G$-orbit of $ q _0 $, $ \mathcal{O} _{ q _0 }=\{ g ^\ast q_0\mid g \in G\}$. In the simplest case, the vector fields $ \xi _i $ only depend on the pointwise values of the tensor fields, denoted as $ \xi _i (q(x),x)$, but more general dependencies are possible, in which for example, the vector fields $ \xi _i $ would depend on the spatial gradients of the tensor fields.

The reduced Eulerian version of this SHP principle reads
\begin{equation}\label{Ham_SVP_2_red}
\delta \int _0^T\Big[\ell(u, g _\ast q_0)+ \big\langle m, dg g ^{-1} - udt - \sum_{i=1}^N \xi _i (  g _\ast q _0 ) \circ d W _i(t) \big\rangle_ \mathfrak{g}  \Big] =0
\,.
\end{equation}
The stochastic Hamilton-Pontryagin principle \eqref{Ham_SVP_2} and its reduced form \eqref{Ham_SVP_2_red} extend to the stochastic case the Hamilton-Pontryagin principles with advection developed in \cite{GBYo2015}. 

By comparing \eqref{Ham_SVP_2_red} with \eqref{Reduced_SVP}, we observe that after the introduction of symmetry breaking in this SHP principle \eqref{Ham_SVP_2} in allowing the stochastic Hamiltonians in \eqref{H_i_2}  to depend functionally on the initial advected quantities $q_0$ the reduction process has allowed the vector fields $ \xi _i $, $i=1,...,N$ in the RSHP \eqref{Ham_SVP_2_red} to depend on the advected tensor field $q(t)= (g_t) _\ast  q_0$, as we have sought.

Applying Hamilton's principle to the reduced action integral in \eqref{Ham_SVP_2_red} now results in the stochastic motion equation for the model that we will discuss in this section, 
\begin{equation}\label{model2} 
d \frac{\delta \ell}{\delta u}+ \operatorname{ad}^*_ {dx_t} \frac{\delta \ell}{\delta u}  = \bigg( \frac{\delta \ell}{\delta q} dt
- \sum_{i=1}^N 
\Big(
\frac{\delta \ell}{\delta u} \cdot \frac{\partial \xi _i }{\partial q}
\Big)  
\circ dW _i (t)
\bigg)  \diamond q 
\,,
\end{equation} 
where $dx_t:=u \,dt+\sum_{i=1}^N \xi _i (q)\circ dW _i (t) $ and the advected tensor field $q(t)= (g_t) _\ast q_0 $ verifies 
\begin{equation}\label{model2-advec} 
d q+ \pounds _{dx _t }q=0
\,. 
\end{equation} 
The expression of the coadjoint operators $ \operatorname{ad}^*$ in the compressible and incompressible cases are recalled in equation \eqref{formula_adstar}. 

In \eqref{model2}, the dot product $(\frac{\delta \ell}{\delta u} \cdot \frac{\partial \xi _i }{\partial q})$ is the composition of two linear maps. The  contraction in the dot product is taken on the vector indices of the variational derivative $\frac{\delta \ell}{\delta u}$ with the derivative of the vector field $ \frac{\partial \xi _i }{\partial q}$  of $\xi _i (q) $ with respect to $q$,  
so that, using \eqref{definition_diamond}, 
\begin{equation}\label{model2-diamond} 
\left\langle \left(\frac{\delta \ell}{\delta u} \cdot \frac{\partial \xi _i }{\partial q}\right)\diamond q
\,,\, v
\right\rangle_\mathfrak{g}
=
\left\langle 
\left(\frac{\delta \ell}{\delta u} \cdot \frac{\partial \xi _i }{\partial q}\right)
\,,\,
\pounds_v q
\right\rangle_V,
\end{equation} 
for $ v \in \mathfrak{g}  $.
For a density, $q=\rho$, we have 
$\langle \phi\diamond\rho\,,\,v \rangle
=
\langle \phi \,,\,{\rm div}(\rho v) \rangle$,
so
$\langle \phi\diamond\rho\,,\,v \rangle
=
- \langle \rho\nabla \phi\,,\,v \rangle$ for a scalar function $\phi$. Hence, in this example, we have
\[
\left\langle \left(
\frac{\delta \ell}{\delta u} \cdot \frac{\partial \xi _i }{\partial \rho}\right) \diamond \rho
\,,\, v
\right\rangle_\mathfrak{g}
=-
\left\langle 
\rho\nabla \left(
\frac{\delta \ell}{\delta u} \cdot \frac{\partial \xi _i }{\partial \rho}\right)
\,,\, v
\right\rangle_\mathfrak{g}
\,.\]

\paragraph{Hamiltonian structure.} From right-invariance of the Hamiltonian \eqref{H_i_2} under the isotropy subgroup $G_{q_0 }$, we obtain the reduced stochastic Hamiltonians
\begin{equation}\label{reduced_hi_def2} 
h _i : \mathfrak{g}  ^\ast\times \mathcal{O} _{q _0 } \rightarrow \mathbb{R}  , \quad h _i ( m , q):= \left\langle m , \xi _i (q) \right\rangle_ \mathfrak{g}  , \quad i=1,..., N \,,
\end{equation} 
see the Appendix for more discussion. 
Upon denoting by $h: \mathfrak{g}  ^\ast \times  \mathcal{O} _{ q _0 }\rightarrow \mathbb{R}  $, the Hamiltonian associated to $\ell$, the above system can be written equivalently as
\begin{equation}\label{syst_h_2}
\left\{
\begin{array}{l}
\displaystyle \vspace{0.2cm}d  m + \operatorname{ad}^*_ {\frac{\delta h}{\delta m }} m \, dt+ \sum_{i=1}^N \operatorname{ad}^*_ {\frac{\delta h _i }{\delta m  } } m  \circ dW _i (t) =-\Big( \frac{\delta h}{\delta q}   dt+\sum_{i=1}^N \frac{\delta h_i }{\delta q}  \circ dW _i (t) \Big)  \diamond q
\,,\\
\displaystyle dq + \pounds _{ \frac{\delta h}{\delta m } }q\,dt
+ \sum_{i=1}^N \pounds _{ \frac{\delta h _i }{\delta m }  }q \circ dW _i (t) =0
\,.
\end{array}
\right.
\end{equation}
One may check that system \eqref{syst_h_2} is Hamiltonian with respect to the Poisson bracket
\begin{equation}\label{Poisson_S4} 
\{f,g\}_{\rm red}( m , q )=  \left\langle m , \left[\frac{\delta f}{\delta m }, \frac{\delta g}{\delta m }\right] \right\rangle_ \mathfrak{g}    + \left\langle  \pounds _{\frac{\delta f}{\delta m }}q  , \frac{\delta g}{\delta q} \right\rangle_V  -  \left\langle \pounds _{\frac{\delta g}{\delta m }}q    , \frac{\delta f}{\delta q} \right\rangle_V\,,
\end{equation}
on $ \mathfrak{g}  ^\ast \times  \mathcal{O}_{q _0 } \ni ( m ,q )$. This Poisson bracket is inherited by Poisson reduction of the canonical Poisson bracket on $T^*G$, by the isotropy subgroup $G_{q _0  } \subset G$ of $q_0 $, namely, the map
\[
(g, \pi ) \in T^*G \mapsto (m, q)= ( \pi g ^{-1} , g _\ast q _0 ) \in (T^*G)/ G_{q_0  }\simeq \mathfrak{g}  ^\ast \times  \mathcal{O}_{q_0 }
\]
is Poisson with respect to these brackets, as discussed in the Appendix. 
Thus, the system \eqref{syst_h_2} admits the Stratonovich-Poisson formulation
\begin{equation}\label{SP_S4} 
df = \{f,h\}_{\rm red} dt+\sum_{i=1}^N \{f, h_i \}_{\rm red}\circ d W _i (t)  =0,
\end{equation} 
where both $h$ and $ h _i $ depend on $m $ and $q$.

\begin{remark}[It\^o forms]\rm
The It\^o forms of the equations in \eqref{syst_h_2} are rather involved. In particular, the It\^o correction in the drift term of the momentum equation is
\begin{equation}\label{Ito_drift1} 
\begin{aligned} 
M_{\rm Ito}dt=&- \frac{1}{2} \left[
 \operatorname{ad}^*_{ \xi _i (q)} \operatorname{ad}^*_{ \xi _i (q)} m 
 +  \operatorname{ad}^*_{ \xi _i (q)} \left( \Big(m \cdot\frac{\partial \xi _i }{\partial q}\Big)\diamond q\right) \right.
\\& \qquad \left.
 +  \Big( \operatorname{ad}^*_{ \xi _i (q)}m \cdot \frac{\partial \xi _i }{\partial q} \Big) \diamond q 
 + \left(  \Big(  (m \cdot\frac{\partial \xi _i }{\partial q}) \diamond q\Big) \cdot  \frac{\partial \xi _i }{\partial q}\right) \diamond q  \right. 
\\
& \qquad \left.
+ \operatorname{ad}^*_{ \frac{\partial \xi _i }{\partial q} \cdot \pounds _{ \xi _i (q)} q}m + m\cdot\frac{\partial ^2\xi_i }{\partial q ^2 } ( \, \cdot \,, \pounds _{ \xi _i (q)} q ) \diamond q + m\cdot \frac{\partial \xi _i }{\partial q} \diamond   \pounds _{ \xi _i (q)} q \right]dt
\end{aligned} 
\end{equation} 
Likewise, the It\^o correction in the drift term of the advection equation is
\begin{equation}\label{Ito_drift2}
A_{\rm Ito}dt=-\frac{1}{2}\left[ \pounds _{ \xi _i (q)} \pounds _{ \xi _i (q)}  q+ \pounds _{ \frac{\partial \xi _i }{\partial q} \cdot \pounds _{\xi _i (q)}q} q \right]dt\, .
\end{equation} 
The drift terms $M_{\rm Ito}dt$ and $A_{\rm Ito}dt$ above follow from the double bracket terms in equation \eqref{Ito_Poisson}. 
\end{remark}

\begin{example}[Rotating shallow water]{\rm Upon choosing the Lagrangian \eqref{l_RSW} for the rotating shallow water equation, we obtain from \eqref{model2} 
\[
d u+ \pounds _ { dx_t}( u +R ) = \nabla\Big( \frac{1}{2} |u | ^2  
+ R \cdot u -g( \eta - B) \Big) dt 
-  \sum_{i=1}^N\nabla \Big( \eta( u + R  )\cdot \frac{\partial \xi  _i }{\partial \eta }   \Big)\circ dW_i (t)\, ,
\]
where $dx_t:=u dt+\sum_{i=1}^N \xi _i (\eta)\circ dW _i (t)$. 
This equation may be written more explicitly as
\begin{equation}\label{Stoch_RSW} 
\begin{aligned} 
&d u + \left( u\cdot \nabla u + \operatorname{curl} R \times  u \right) dt+\sum_{i=1}^N\pounds _{  \xi _i } (u + R ) \circ d W _i (t)\\
&\qquad \qquad \qquad\qquad\qquad =-g\nabla ( \eta - B)dt
-\sum_{i=1}^N  
\nabla \Big( \eta( u + R  )\cdot \frac{\partial \xi  _i }{\partial \eta }   \Big)
  \circ dW_i (t) \,.
\end{aligned}
\end{equation} 
The advection equation for the surface elevation in this model is 
\begin{equation}\label{advection_SW} 
d\eta + \operatorname{div}( \eta dx _t )=0 
\,.
\end{equation} 
Taking the curl of the momentum equation yields the corresponding RSW vorticity equation  
\[
d\omega + {\rm div}(\omega d x_t) = 0
\quad\hbox{for}\quad
\omega={\rm curl}\left(\frac{1}{\eta} \frac{\delta \ell}{\delta u }\right) \cdot \mathbf{\hat{z}}  = {\rm curl}(u+ R) \cdot \mathbf{\hat{z}} 
\,.\]
This, together with the advection equation \eqref{advection_SW} yields the potential vorticity (PV) equation,
\[
dQ + d x_t \cdot \nabla Q = 0
\quad\hbox{for}\quad
Q := \omega/\eta
\,,
\]
which expresses conservation of potential vorticity along the stochastic fluid particle path $d x_t $. 
}
\end{example}

\medskip

\begin{remark}{\rm
We shall now consider a specific expression for $ \xi _i (h)$, that yields a simplified expression for the stochastic PDE \eqref{Stoch_RSW}. First, we note that by assembling the stochastic terms in \eqref{Stoch_RSW}, we get, in slightly abbreviated notation,
\begin{align*} 
&\pounds _{  \xi _i } (u + R ) 
+  \nabla \Big( \eta( u + R  )\cdot \frac{\partial \xi  _i }{\partial \eta }   \Big)
= \xi _i \cdot \nabla (u+R)+ \nabla \xi _i ^\mathsf{T} \cdot (u+R) 
+  \nabla \Big( \eta( u + R  )\cdot \frac{\partial \xi  _i }{\partial \eta }   \Big)
\\
&\quad = \xi _i \cdot \nabla (u+R)+ \nabla (\xi _i \cdot (u+R))- \nabla (u+R)^\mathsf{T}\cdot \xi _i  
+  \nabla \Big( \eta( u + R  )\cdot \frac{\partial \xi  _i }{\partial \eta }   \Big)
\\
&\quad = \operatorname{curl}(u+R)\times\xi _i 
+  \nabla \Big( \xi _i \cdot (u+R)
+   \eta( u + R  )\cdot \frac{\partial \xi  _i }{\partial \eta }   \Big).
\end{align*} 
Upon making the choice $\xi_i (\eta):= -  \eta^{-1}X_i$, for a set of \textit{fixed vector fields} $X _i $, $i=1,...,N$, we have 
$\frac{\partial \xi_i }{\partial \eta}=  \eta^{-2}X_i$ and 
$ \frac{\partial \xi _i }{\partial \eta} ^\ast \cdot m= \eta^{-2} (m\cdot X _i )$, so that
\[
\nabla \Big( \xi _i \cdot (u+R)
+   \eta( u + R  )\cdot \frac{\partial \xi  _i }{\partial \eta }   \Big)=0
\,,
\]
in the expression above. In this case, the stochastic RSW equations \eqref{Stoch_RSW} take the simpler form
\begin{equation}\label{RSW_simpler} 
d u + \left( u\cdot \nabla u + \operatorname{curl} R \times  u \right) dt=-g\nabla ( \eta - B)dt-\sum_{i=1}^N\operatorname{curl}(u+R)\times\xi _i (\eta)  \circ dW_i (t) .
\end{equation} 
Having chosen the simpler form above with $\xi_i (\eta):= -  \eta^{-1}X_i $, we may write the stochastic term of \eqref{RSW_simpler} in terms of the potential vorticity $Q$ as
\[
\operatorname{curl}(u+R)\times\xi _i (\eta)  \circ dW_i (t)= -\, Q\, X_i ^\perp\circ dW_i (t).
\]}
\end{remark}

\medskip

\begin{example} [Rotating compressible barotropic fluid]{\rm Upon choosing the tensor field $q(t,x)$ to be the mass density $ \rho(t,x) $, the stochastic model \eqref{model2} yields
\[
d \frac{\delta \ell}{\delta u}+\pounds _ {dx_t} \frac{\delta \ell}{\delta u}  
=  \rho \nabla \Big( \frac{\delta \ell}{\delta \rho } dt
- \sum_{i=1}^N 
\Big( \frac{\partial \xi _i }{\partial\rho } \cdot  \frac{\delta \ell}{\delta u}\Big)  \circ dW _i (t)\Big),
\]
where $dx_t:=u dt+\sum_{i=1}^N \xi _i (\rho )\circ dW _i (t) $.
Upon using the advection equation,
\[
d\rho + \operatorname{div}( \rho\,  dx _t )=0\,,
\]
the previous equation can be written equivalently as
\[
d \left( \frac{1}{ \rho }\frac{\delta \ell}{\delta u} \right) 
+\pounds _ {dx_t}  \frac{1}{ \rho }\frac{\delta \ell}{\delta u}  
=   \nabla \Big( \frac{\delta \ell}{\delta \rho } dt
- \sum_{i=1}^N 
\left( \frac{\partial \xi _i }{\partial\rho } \cdot \frac{\delta \ell}{\delta u} \Big) \circ dW _i (t)\right).
\]
Upon taking the Lagrangian \eqref{Lagr_comp} for the compressible barotropic fluid, we obtain the stochastic motion equation,
\begin{align*} 
du+ \Big( u \cdot \nabla u+ \operatorname{curl} R \times u +   g \mathbf{z} + \frac{1}{ \rho }\nabla p  \Big) dt&=-\sum_{i=1}^N \Big( \operatorname{curl}(u+R) \times \xi _i +\nabla ( \xi_i \cdot (u+R)) \Big)\circ dW _i (t)\\
&  \qquad \qquad 
- \sum_{i=1}^N \nabla \left(  \frac{\partial \xi _i }{\partial \rho } \cdot\rho (u+R)  \right) \circ d W _i(t),
\end{align*} 
in which the thermodynamic pressure is defined by $p=\rho ^2 \frac{\partial e}{\partial\rho }$.

\medskip

Taking the curl of the momentum equation and using the advection equation $d\rho + \operatorname{div}( \rho\,  dx _t )=0$ now implies PV conservation as
\[
dQ + dx _t \cdot \nabla Q=0, \quad \text{where} \quad  Q:=  \frac{1}{\rho } \operatorname{curl}\left(  \frac{1}{ \rho } \frac{\delta \ell}{\delta u}\right)  \cdot\nabla \rho = \frac{1}{\rho } \operatorname{curl}(u+R) \cdot\nabla \rho 
\,.
\]
Similarly to the case of the RSW equations, with the choice $\xi _i (\rho )= -\,X _i/\rho$
for a set of fixed vector fields $X _i$, $i=1,...,N$, the previous motion equation simplifies to
\[
du+ \Big( u \cdot \nabla u+ \operatorname{curl} R \times u +   g \mathbf{z} + \frac{1}{ \rho }\nabla p  \Big) dt=-\sum_{i=1}^N  \operatorname{curl}(u+R) \times \xi _i (\rho)   \circ dW _i (t).
\]

\begin{remark} \rm
The last two examples preserve the integrals
\[
C_\Phi = \int _ \mathcal{D} \rho \,\Phi(Q)\,d^3x,
\]
for smooth functions $\Phi $, as may be seen either from the Hamiltonian formulation \eqref{SP_S4} with Poisson bracket \eqref{Poisson_S4} with $q= \rho $, or by direct verification. 
{These conserved quantities are Casimirs for the Poisson bracket. They are the same as the conserved potential vorticity integrals in the deterministic case, because the Poisson bracket persists in passing to the stochastic case.}
\end{remark}}
\end{example}

\section{Conclusions} \label{sec-conclus}
This paper has used standard methods from symmetry breaking in geometric continuum mechanics, as discussed for example in \cite{HoMaRa1998}, \cite{GBTr2010}, to set out three different approaches for incorporating stochastic transport into ideal fluid dynamics. As mentioned in the Introduction, our approach preserves the relabelling symmetry of fluid dynamics, up to isotropy of the advected quantities. Therefore, we may summarise the results of these approaches, simply by comparing their advection laws and the symmetry breaking terms of their respective Kelvin-Noether circulation theorems.

\paragraph{Model 1 -- Time independent spatial correlation eigenfunctions.} In Model 1, reviewed in section \ref{sec-Holm2015Review}, the time-independent spatial statistical correlation eigenvectors $\xi_i(x)$ are obtained as eigenvectors of an appropriate correlation function, which is assumed to be time-independent. From equations \eqref{LP_stoch} and \eqref{ed_hi}, possibly extended to include advected quantities $q$, the corresponding Kelvin circulation theorem is given by
\begin{align}
d\oint_{c(dx_t)}\frac{1}{ \rho }\frac{\delta \ell}{\delta u} \cdot dx
=
\oint_{c(dx_t)} \Big(\frac{1}{ \rho }\frac{\delta \ell}{\delta q}  \diamond q \,dt \Big)\cdot dx
\,,\label{KelThm-Mod1}
\end{align}
where $ \rho $ is the mass density obeying the continuity equation, $d\rho + \operatorname{div}( \rho dx_t)=0$, and the advected tensor field $q= (g_t )_* q_0$ satisfies 
\begin{align}
d q+ \pounds_{dx_t }q  =0\,,
\label{advec-Mod3}
\end{align}
where $\pounds_v q$ denotes Lie derivative of advected quantity $a$ by the stochastic vector field  
\[
dx_t=u(x,t)dt + \sum_{i=1}^N\xi_i(x)\circ d W _i(t)
\,.
\]
This is the model introduced in \cite{Ho2015}. 

\paragraph{Model 2 -- Time-dependent advected statistical correlation eigenvectors.} From equation \eqref{red_SVP_3_advect-relat} for Model 2, discussed in section \ref{sec_advection} we have the circulation dynamics
\begin{align}
d\oint_{c(u)} \frac{1}{ \rho }\frac{\delta \ell}{\delta u} \cdot dx
=
\oint_{c(u)} \Big(\frac{1}{ \rho }\frac{\delta \ell}{\delta q}  \diamond q \,dt \Big)\cdot dx
+
\oint_{c(u)} \Big(\frac{1}{ \rho }\sum_{i=1}^N \operatorname{ad}^*_ { \zeta _i } \frac{\delta h _i }{\delta \zeta _i }  \circ dW _i (t)\Big)  \cdot dx
\,,
\label{KelThm-Mod2}
\end{align}
and  with ${\delta h _i }/{\delta m} = 0$ the advected tensor fields $q (t) = (g_t ) _\ast q_0 $ and $\zeta_i (t) = {\rm Ad}_{g_t}\xi_i$ verify
\begin{align}
d q+ \pounds_{u}q \,dt =0\,,
\quad\hbox{and}\quad
d \zeta _i + [\zeta_i,u]dt=0\,.
\label{advec-Mod3}
\end{align}
In particular, the mass density verifies the ordinary continuity equation $ d \rho + \operatorname{div}(\rho u) dt=0$. 

This means that in Model 2, treated in section \ref{sec_advection}, the choice of stochastic Hamiltonians satisfying $\delta h _i /\delta m=0$ implies that  the transport velocity of the circulation loop in \eqref{KelThm-Mod2} reduces to the drift velocity, $u(x,t)dt$, alone, rather than the entire stochastic velocity $dx_t$, as in Model 1. Moreover, for $\delta h _i /\delta m=0$, the Stratonovich and It\^o representations of Model 2 take the same form.

\begin{remark}{\rm
In deriving the Kelvin theorem in equation \eqref{KelThm-Mod2} from \eqref{red_SVP_3_advect-relat}, we have used the relation
\begin{align}
\sum_{i=1}^N \operatorname{ad}^*_ { \zeta _i } \frac{\delta h _i }{\delta \zeta _i } = \sum_{i=1}^N \frac{\delta h _i }{\delta \zeta _i } \diamond \zeta _i\,,
\label{adstar-diamond-equiv}
\end{align}
which arises from equivalence of the definitions of ${\rm ad}^\ast$ and the diamond operator $(\diamond)$ when the advected quantity is a vector field. This notation makes it clear that the drift term and the stochastic  term on the right-hand side of equation \eqref{KelThm-Mod2} both arise from the same source; namely, symmetry breaking of the Lie group of diffeomorphisms to the isotropy subgroup of the initial condition for an advected quantity.}
\end{remark}

\paragraph{Model 3 -- Time-dependent spatial correlation eigenvectors depending on advected quantities.} From equation \eqref{model2} for Model 3, discussed in section \ref{sec-evsadvectedquantities}, we have the circulation dynamics 
\begin{align}
d \oint_{c(dx_t)}  \frac{1}{\rho } \frac{\delta \ell}{\delta u} \cdot dx
= 
\oint_{c(dx_t)} \frac{1}{ \rho } 
\left[ \frac{\delta \ell}{\delta q} dt
- \sum_{i=1}^N 
\left(\frac{\delta \ell}{\delta u} \cdot \frac{\partial \xi _i }{\partial q}\right) \circ dW _i (t)\right]  \diamond q  \cdot dx
\,,
\label{KelThm-Mod3}
\end{align}
where $dx_t:=u dt+\sum_{i=1}^N \xi _i (q)\circ dW _i (t) $ and the advected tensor fields $q= (g_t) _\ast q _0$ verify
\begin{align}
d q+ \pounds_{dx _t}q  =0\,,
\label{advec-Mod3}
\end{align}
where $\pounds_{dx_t} q$ denotes Lie derivative of advected quantity $q$ by the stochastic vector field  
\[
dx_t:=u dt+\sum_{i=1}^N \xi _i (q)\circ dW _i (t)
\,.\]
In particular the mass density satosfies $ d \rho + \operatorname{div}(\rho dx_t)=0$.  

In Model 3,  treated in section \ref{sec-evsadvectedquantities}, the velocity of the circulation loop is $dx_t$. As in Model 1 this is the full stochastic velocity, except now its spatial correlation eigenvectors $\xi _i (q)$ respond to the motion of the advected quantities which also contribute to the thermodynamic and potential energy properties of the fluid. 

\medskip

\begin{remark}\rm 
The It\^o forms of these three models have been computed in the body of the paper, although the additional It\^o drift terms may take a more complicated form than in the Stratonovich case. An example is the additional It\^o drift term $M_{\rm Ito}dt$ in \eqref{Ito_drift1} for Model 3. Nonetheless, the equivalent It\^o forms of their circulation theorems may be computed by the standard methods from their Stratonovich forms; namely, by transforming both the integrand and the velocity of the loop into It\^o form, using the corresponding motion equations.
\end{remark}

\begin{remark}\rm
Besides the initial and boundary conditions, the key to applying the methods of this paper efficiently begins with the choice of the spatial correlation eigenvectors. Once a data assimilation step for determining the eigenvectors is completed, such as the one discussed in Remark \ref{corr-eofs} based on \cite{CoCrHoPaSh2017}, then the stochastic Hamilton's principle provides the dynamics of the drift velocity. Thus, the present approach may be regarded as a data driven approach to dynamically and self-consistently separating the Lagrangian paths into their drift dynamics and stochastic parts. This question needs to be confronted in any application to any data set. Model 1 chooses to model the stochastic parts by using optimal eigenvectors of an equilibrium spatially-constant time-independent correlation function. Models 2 and 3 provide two different approaches for making these correlation eigenvectors time-dependent. For a detailed discussion of methods for modelling the separation between drift and stochasticity in comparing numerical simulations of fluid flows at fine and coarse resolutions, see \cite{CoCrHoPaSh2017}.

In applications, these three models of the stochastic fluid velocity vector field could be used either separately, or in combinations, by taking combinations of the correlation eigenvectors to evolve in any way that might be needed for tuning the stochastic transport, e.g., in applying them for data assimilation. 

In conclusion, we mention that the three models of stochastic fluid dynamics treated here may also be transferred into data assimilation methods in biomedical image analysis for computational anatomy. In particular, the approach of \cite{ArHoPaSo2017} for stochastic image analysis based on Model 1 could be extended to create new approaches to computational anatomy based on the ideas underlying Model 2 and Model 3. 

\end{remark}

\section*{Appendix}

The Appendix provides additional background for some of the standard notations from geometric mechanics used in the paper and gives some details on the structure of the reduced Poisson brackets $\{\,\cdot\,,\cdot\,\}_{\rm red}$ arising in Model 2 and Model 3. For further geometric mechanics background, see \cite{MaRa1994, Ho2011}.

\paragraph{Lie group notation.} Given a Lie group $G$, and two elements $g,h \in G$, we denote by $gh$ the Lie group multiplication. The spaces $T_gG$ and $T^*_gG$ refer to the tangent and cotangent spaces of $G$ at $g$. For $v \in T_gG$ and $ \pi \in T^*_gG$, we denote by $vh \in T_{gh}G$ and $\pi h \in T^*_{gh}G$, the right translation of $ v$ and $\pi $ by $h \in G$. These right translations are defined as follows:
\[
vh:= \left.\frac{d}{d\varepsilon}\right|_{\varepsilon=0} g_ \varepsilon h \qquad\text{and}\qquad \left\langle \pi h, v \right\rangle :=\left\langle \pi , v h ^{-1} \right\rangle , \;\; \text{for all $v \in T_{gh}G$}, 
\]
where $g_ \varepsilon \in G$ is a path with $g_{ \varepsilon =0}=g$ and $ \left.\frac{d}{d\varepsilon}\right|_{\varepsilon=0} g_ \varepsilon =v$, and $ \left\langle \,\cdot\,,\cdot\,\right\rangle $ is the duality pairing between the tangent and cotangent spaces of $G$. We have $\mathfrak{g}  =T_eG$ and $ \mathfrak{g}  ^\ast = T_e ^\ast G$, where $e$ is the neutral element in $G$. The adjoint action of $g \in G$ on $u \in \mathfrak{g}  $ is defined by
\[
\operatorname{Ad}_g u:=\left.\frac{d}{d\varepsilon}\right|_{\varepsilon=0} g h_ \varepsilon g ^{-1} , 
\]
where $h_ \varepsilon \in G$ is a path with $ h_{ \varepsilon =0}=e$ and $ \left.\frac{d}{d\varepsilon}\right|_{\varepsilon=0} h_ \varepsilon =u$.

When $G$ is the group of diffeomorphism of the fluid domain $ \mathcal{D} $, the multiplication $gh$ corresponds to the composition of diffeomorphisms. Elements $v \in T_gG$ are vector field on $ \mathcal{D} $ covering the diffeomorphism $g$ and $vh$ is the composition on the right by the diffeomorphism $h$. In particular, elements in the Lie algebra $ \mathfrak{g}  $ are vector fields on $ \mathcal{D} $. In the paper, we have extensively used the fact that on groups of diffeomorphisms, $ \operatorname{Ad}_g u  = g _\ast u$, the push-forward of vector fields.

\paragraph{Reduced Poisson brackets.} The Poisson brackets \eqref{PB_S3} and \eqref{Poisson_S4} of models 2 and 3 both have the same structure, which we now explain. Both brackets are inherited from the canonical Poisson bracket $\{\,\cdot\,,\cdot\,\}_{\rm can}$ on $T^*G$, by a reduction by symmetry associated to a subgroup of $G$. 

Consider a right action of $G$ on a manifold $P$, denoted as follows $(n,g) \in P \times G \mapsto  \phi _g(n) \in P$. Given $n_0 \in P$, we denote by $G_{n_0}=\{ g \in G\mid \phi _g (n_0)=n_0\} \subset G$ the isotropy group of $n_0$ and by $ \mathcal{O} _{n_0}=\{ \phi _g(n)\mid g \in G\} \subset P$ the orbit of $n_0$. The Lie algebra action of $u \in \mathfrak{g}  $ on $n \in N$ is written as
\[
n\! \cdot\! u:= \left.\frac{d}{d\varepsilon}\right|_{\varepsilon=0}\phi _{g_ \varepsilon }(n)\in T_nN,
\]
were $g_ \varepsilon\in G$ is a path with $g_{ \varepsilon =0}=e$ and $ \left.\frac{d}{d\varepsilon}\right|_{\varepsilon=0} g _\varepsilon =u$.

The quotient space of $T^*G$ by $G_{n_0}$ relative to the action  given by $(g, \pi ) \in T^*G\mapsto ( gh , \pi h) \in T^*G$, $h \in G_{n_0}$, is denoted as $(T^*G)/G_{n_0}$. One observes that we have the identification $(T^*G)/G_{n_0}\simeq \mathfrak{g}  ^\ast \times  \mathcal{O} _{n_0}$ given by 
\[
[g, \pi ]_{G_{n_0}} \in (T^*G)/G_{n_0} \mapsto (m, n)= \left(  \pi g ^{-1} ,\phi _{g ^{-1} }(n_0)\right)  \in \mathfrak{g}  ^\ast \times  \mathcal{O} _{q_0}.
\]
A $G_{n_0}$-invariant Hamiltonian $H:T^*G \rightarrow \mathbb{R}  $ thus induces the reduced Hamiltonian $h: \mathfrak{g}  ^\ast \times  \mathcal{O} _{n_0} \rightarrow \mathbb{R}  $ defined by \begin{equation}\label{red_Hamiltonian} 
H( g, \pi )= h \left(  \pi g ^{-1} ,\phi _{g ^{-1} }(n_0) \right).
\end{equation}
This formula has been used several times in the paper, for instance in \eqref{reduced_hi_def} and \eqref{reduced_hi_def2} for the stochastic Hamiltonians.
The reduced Poisson bracket on $ \mathfrak{g}  ^\ast\times \mathcal{O} _{ n_0}$ induced by the canonical Poisson bracket $\{\,\cdot\,,\cdot\,\}_{\rm can}$ on $T^*G$ is given by
\begin{equation}\label{red_PB}
\{f,g\}_{\rm red}(m, n)= \left\langle m, \left[ \frac{\delta f}{\delta m},\frac{\delta g}{\delta m}  \right] \right\rangle_ \mathfrak{g}   + \left\langle \frac{\delta h}{\delta n}, n\!\cdot\! \frac{\delta f}{\delta m} \right\rangle-\left\langle \frac{\delta f}{\delta n}, n\!\cdot \!\frac{\delta g}{\delta m} \right\rangle,
\end{equation} 
for functions $f,g:  \mathfrak{g}  ^\ast\times \mathcal{O} _{ n_0} \rightarrow \mathbb{R}$, see, e.g,   \cite{GBTr2010} (Theorem 2.7) for a proof of this fact. The last two terms involve the pairing $ \left\langle \,\cdot\,,\cdot\,\right\rangle $ between $T ^\ast  \mathcal{O} _{n_0}$ and $T \mathcal{O} _{n_0}$. We now explain how this Poisson reduction $T^*G \rightarrow \mathfrak{g}  ^\ast\times \mathcal{O} _{ n_0}$ applies to Models 2 and 3.

For Model 2, we have $P= \mathfrak{g}  \times ... \times\mathfrak{g}\ni ( \zeta _1 , ..., \zeta _N ) $, $n_0= ( \xi  _1 , ..., \xi  _N )= \xi $ and the action $ \phi _g$ given by the push-forward of vector fields $ \operatorname{Ad}_g \xi = g _\ast \xi $ on each term. With these choices, the reduced Poisson bracket \eqref{red_PB} recovers the expression \eqref{PB_S3}. For the system \eqref{red_SVP_3_advect-relat}, we take $P= V \times \mathfrak{g}  \times ... \times\mathfrak{g}\ni ( q, \zeta _1 , ..., \zeta _N ) $, $n_0=( q_0, \xi _1 , ..., \xi _N )$ and the action $ \phi _g$ given by the push-forward action on each terms.

For Model 3, we have $P=V\ni q$, $n_0= q _0 $ and the action $ \phi _g $ given by the push-forward of tensor fields, $ \phi _g(q)= g^\ast q$. With these choices, the reduced Poisson bracket \eqref{red_PB} recovers the expression \eqref{Poisson_S4}.

\paragraph{Acknowledgements.} We are very grateful to C. J. Cotter, D. Crisan, E. M\'emin,  S. Olhede, W. Pan, V. Resseguier,  I. Shevshenko, and A. Sykulski for valuable discussions and presentations during the course of this work. During this work, FGB was partially supported by the ANR project GEOMFLUID 14-CE23-0002-01 and DDH was partially supported by the European Research Council Advanced Grant 267382 FCCA and EPSRC Standard Grant EP/N023781/1.


\end{document}